\documentclass[fleqn,usenatbib]{class}

\usepackage{newtxtext,newtxmath}
\usepackage{float}
\restylefloat{figure}
\restylefloat{table}
\usepackage{tabularx}
\usepackage[toc,page]{appendix}

\usepackage[paper=a4paper,left=17mm,right=17mm,top=20mm,bottom=20mm]{geometry}

\usepackage{caption}
\usepackage{setspace}
\usepackage{comment}
\usepackage{natbib}
\usepackage{graphicx, animate}

\usepackage{color,soul}

\usepackage{hyperref}
\hypersetup{
    colorlinks=true,
    linkcolor=ultramarine3,
    filecolor=ultramarine3,      
    urlcolor=ultramarine3,
    pdftitle=Weinberg angle and fine-structure constant,
    pdfpagemode=FullScreen,
    citecolor=ultramarine3
    }

\usepackage{xcolor}
\definecolor{ultramarine}{RGB}{0,32,96}
\definecolor{ultramarine2}{RGB}{0,68,204}
\definecolor{ultramarine3}{RGB}{0,0,180}

\usepackage[T1]{fontenc}

\DeclareRobustCommand{\VAN}[3]{#2}
\let\VANthebibliography\thebibliography
\def\thebibliography{\DeclareRobustCommand{\VAN}[3]{##3}\VANthebibliography}

\usepackage{graphicx,subfigure}	
\usepackage{amsmath}	
\usepackage{amsmath,array,graphicx}
\usepackage{kantlipsum}
\usepackage{hyperref}

\newcommand{\eqb}{\begin{equation}}
\newcommand{\eqe}{\end{equation}}
\newcommand{\dmb}{\begin{displaymath}}
\newcommand{\dme}{\end{displaymath}}

\newcommand{\eab}{\begin{eqnarray}}
\newcommand{\eae}{\end{eqnarray}}

\newcommand{\be}{\begin{equation}}
\newcommand{\ee}{\end{equation}}

\title
[$\alpha, \theta_W$ from YM Thermodynamics]
{Electroweak parameters from mixed SU(2) Yang-Mills Thermodynamics}

\author[Meinert and Hofmann]{
Janning Meinert$^{2,1}$\thanks{E-mail: J.Meinert@ThPhys.Uni-Heidelberg.de}\href{https://orcid.org/0000-0001-7582-3456}{\hspace{0.1mm}\includegraphics[scale=0.06]{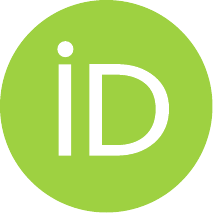}},
Ralf Hofmann$^{1}$\thanks{
E-mail: R.Hofmann@ThPhys.Uni-Heidelberg.de \newline
Both authors contributed equally to this work.
}\href{https://orcid.org/0000-0001-6365-0631}{\hspace{0.1mm}\includegraphics[scale=0.06]{orcid.pdf}
}
\vspace{2mm}\\
$^{1}$Institut f\"ur Theoretische Physik, Universit\"at Heidelberg, Philosophenweg 12, D-69120 Heidelberg, Germany\vspace{0.5mm}\\
$^{2}$Institut f\"ur Physik, Bergische Universit\"at Wuppertal, Gaußstraße 20, D-42119 Wuppertal, Germany\vspace{0.5mm}\\
}

\begin{document}
\label{firstpage}
\pagerange{\pageref{firstpage}--\pageref{lastpage}}
\maketitle


\begin{abstract}
\noindent Based on the thermal phase structure of pure SU(2) quantum Yang-Mills theory, we describe the electron at rest as an extended particle, a droplet of radius $r_0\sim a_0$, where $a_0$ is the Bohr radius.
This droplet is of vanishing pressure and traps a monopole within its bulk at a temperature of $T_c=7.95$\,keV. 
The monopole is the Bogomolny-Prasad-Sommerfield (BPS) limit. It is interpreted in an electric-magnetically dual way. Utilizing a spherical mirror-charge construction, we approximate the droplet's charge at a value of the electromagnetic fine-structure constant $\alpha$ of $\alpha^{-1}\sim 134$ for soft external probes. It is shown that the droplet does not exhibit an electric dipole or quadrupole moment due to 
averages of its far-field electric potential over monopole positions. We also calculate the mixing angle $\theta_{\rm W}\sim 30^{\circ}$ which belongs to deconfining phases of two SU(2) gauge theories of very distinct Yang-Mills scales ($\Lambda_{\rm e}=3.6\,$keV and $\Lambda_{\rm CMB}\sim 10^{-4}\,$eV). Here, the condition that the droplet's bulk thermodynamics is stable determines the value of $\theta_W$. The core radius of the monopole, whose inverse equals the droplet's mass in natural units, is about 1\hspace{0.25pt}\% of $r_0$. 
\end{abstract}

\begin{keywords}
SU(2) Yang-Mills thermodynamics, BPS monopole, fine-structure constant, electroweak mixing angle
\end{keywords}



\section{Introduction\label{sec1}}

The electroweak interactions between leptons in the Standard Model of Particle Physics (SM) are mediated by quantised gauge fields, associated with the group U(1)$_Y\times$SU(2)$_{\rm W}$. In the SM, asymptotic lepton states are associated with point particles whose localisation characteristics relate to quantum states of relativistic matter waves, described by solutions to the Dirac equation, and which depend on the presence or absence of external potentials. Electroweak interactions are assumed to be governed by small gauge couplings: $g^\prime<1$ for U(1)$_Y$ and $g<1$ for SU(2)$_{\rm W}$. The SM is extremely successful and efficient in delivering a quantitative description of scattering cross-sections, decay rates, branching ratios, production and oscillation rates, as well as bound-state properties. Renormalisability of the electroweak sector of the SM \citep{tHooft:1971qjg,tHooft:1971akt,tHooft:1972tcz} assures that corrections to an asymptotic state are calculable to any desired order in the couplings $g^\prime$ and $g$ without having to introduce parameters in addition to those of the defining Lagrangian. By resummation of renormalised Feynman diagrams, these parameters evolve with the resolution scale externally applied to a physical process. Generally speaking, a parameter value obtained at a given four-momentum transfer $P$ can be converted to that at a different four-momentum transfer $P^\prime$ by a rescaling that depends on logarithms of the ratio $P/P^\prime$.\\ 

While the electroweak sector of the SM reliably predicts the running of its parameters and the amplitudes of associated processes, it is unable to compute {\sl absolute} parameter values at any given resolution. Most prominently, the electromagnetic fine-structure constant $\alpha$ \citep{https://doi.org/10.1002/andp.19163561702}, which at low four-momentum transfer assumes the experimental value \citep{alphaNature2020Morel}, 
\begin{equation}
\alpha_{{\rm exp}}=\frac{q^2}{4\pi\epsilon_0 \hbar c}=\frac{1}{137.035999206}\,,
\label{definition_alpha}
\end{equation}
is not subject to calculation within the SM. That is, given $\epsilon_0$ (permittivity of the vacuum), $\hbar$ (reduced quantum of action), and $c$ (speed of light in vacuum), the electric charge of the electron $q$ is a free parameter whose value needs to be measured. In what follows, we assume that the constants of nature $\epsilon_0$, $c$, and $\hbar$ are independently measured and thus are not considered free parameters of the SM\footnote{They can be argued to emerge from pure SU(2) Yang-Mills theory, see \citep{bookHofmann}.}. These constants and Boltzmann's constant $k_{\rm B}$ are all set equal to unity in the present report (natural units). 

In the electroweak sector of the SM \citep{Glashow:1959wxa,Salam:1964ry,Salam:1959zz} the electric charge $q$, due to mixing between the Cartan-algebras of SU(2)$_{\rm W}$ and U(1)$_Y$, is obtained from the (perturbative) coupling $g$ of SU(2)$_{\rm W}$ or from the (perturbative) coupling $g^\prime$ of U(1)$_Y$ as 
\eqb
\label{su2}
q=g\sin\theta_{\rm W} \hspace{0.7cm}\text{or}\hspace{0.7cm}q=g^\prime\cos\theta_{\rm W}\,.
\eqe
Here, $28.7^\circ\le \theta_{\rm W}\le 29.3^\circ$ \citep{PDG22,MoellerscatteringWA} denotes the weak mixing (or Weinberg) angle, measured at variable four-momentum transfer. In addition to condition (\ref{su2}), one imposes for the electric charge operator $Q$ and the hypercharge matrix $Y$ of each left-handed lepton family that $Y=Q-\frac12\sigma_3$ where $\sigma_3={\rm diag}(1,-1)$ denotes the third Pauli matrix, see \citep{pich1994standard} for a comprehensive review. Like $q$, $\theta_{\rm W}$ is a free parameter of the SM whose value has to be inferred experimentally at a given resolution. One can pick resolution to be vanishing, and we argue in the present work that, under clearly stated assumptions, in this limit the values of the dimensionless electroweak parameters $\alpha$ and $\theta_{\rm W}$ can be approximated thermodynamically. { Such an approach is based on ideas by Louis de Broglie providing the foundations for wave mechanics \citep{DeBroglieRev,DeBroglieFP}. 
Namely, the electron as a propagating matter wave emerges by applying a Lorentz boost to a spatially extended and internally oscillating system (a standing wave). De Broglie also noticed that the transformations of wave frequency and the frequency of internal oscillation are distinct in the same way as the transformations of particle energy and internal heat are. This, however, reveals the existence of a close link between relativistic thermodynamics and the physical insights that are at the heart of wave mechanics.\\

The present work takes up these profound ideas by de Broglie, supplemented by concrete notions 
of what the source of oscillation, the oscillating medium, and the boundary of this medium 
are in the framework of pure SU(2) Yang-Mills thermodynamics in four spacetime dimensions. 
As we shall argue, this enables derivations of the values of the electroweak parameters $\alpha$ and $\theta_W$.} 
{ To put this into perspective, we would also like to mention other 
approaches. The experimental value of $\alpha$ in Eq.\,(\ref{definition_alpha}) is 
either represented by numerology or linked to physics beyond that of the SM. Pure numerology without a defined physical basis can generate an impressive proximity to the value in Eq.\,(\ref{definition_alpha}), 
mostly invoking combinations of primes, and prominent transcendental numbers in elementary functions or continued fractions. These results are reviewed in \citep{alphanumerology}. Other approaches, including Dirac’s large number hypothesis as well as Weyl’s hypothesis, both invoking the classical radius of the electron, Casimir’s mousetrap model, Kaluza-Klein theories, open-string scattering, invariants under certain ad hoc symmetries and others are reviewed in \citep{Jentschura_2014}. 
Regarding the electroweak mixing angle $\theta_W$, the assumption that quarks have integer electric charges as in the Han-Nambu scheme solves the orthogonality condition between the $Z$-boson field and the photon-field $\gamma$ by a value $\sin^2\theta_W=1/4$ or $\theta_W=30^\circ$ which is close to experiment \citep{faessler2013weinbergangleintegerelectric}. For discussions on how the Weinberg angle relates to consistency conditions in breaking the symmetry groups of grand unified theories (GUT) in four or higher
dimensions, see \citep{ENGLERT2020135548} and references therein.}\\  

The strength of the SM is that parameters such as $\alpha$ and $\theta_W$ can be evolved accurately to describe particle transitions that are characterised by large four-momentum transfers. To such processes, thermodynamics is not applicable. { This is already suggested by the fact that in the center-of-mass (COM) frame of two colliding 
electrons the square of COM energy is given as $s=4\,m_{e}^2(1-v^2)^{-1}$ and that the temperature $T_{0}$ in each particle's restframe is, by invariance of local entropy under Lorentz boosts \citep{DeBroglieFP}, decreased as $T=\sqrt{1-v^2}T_{0}$. Here, $\pm v$ are the respective particle speeds. Letting $v\nearrow 1$, we notice that, as $s\to\infty$, the temperature of each droplet vanishes, $T\searrow 0$. This associates a strongly boosted electron with the confining phase of SU(2) Yang-Mills thermodynamics by depriving it of its 3D structure. Alternatively, an observer in the center-of-mass frame  
instead of seeing colliding, spherical droplets sees 2D 'pancakes' due to a strong Lorentz contraction along the boosted coordinate. Therefore, the deconfining phase, whose restframe thermodynamics mainly constitutes the droplet and defines its Lorentz invariant properties mass $m_e$ and electric charge $q$, is invisible in the center-of-mass frame at large $s$, see Secs.\,4 and 5.}  \\  

{ The process of pair creation of an electron and a positron, both separated well at late times,
by a pair of initially well separated photons is 
another example of large four-momentum transfer during the transition.

Here, a highly nonthermal, intermediate state of even locally inhomogeneous energy density is generated. This intermediate state emerges from local energy deposition into the ground state of the confining phase of an SU(2) Yang-Mills theory.
The round-point center vortex loops composing this ground state are stretched and twisted by such an energy composition \citep{bookHofmann}. 
Subsequently, this intermediate state equilibrates into a final state, the electron. 
In the restframe of the ensuing electron, a thermal droplet at rest is well separated spatially from the boosted thermal droplet -- the positron. 
In Quantum Electrodynamics (QED) and to lowest order in $\alpha$ the nonthermal, intermediate of a pair creation process, where both particles are not fully developed and close by, is described by the electron propagator connecting the two vertices of the associated Feynman diagram. This propagator mediates off-shell propagation of positive and negative energy states forward and backward through limited time intervals, and 
it is well known how effective and precise 
the associated QED cross sections describe the experimental data on pair creation \citep{PhysRev.76.749}. Therefore, QED (and more generally the SM) is a powerful tool to quantitatively assess collisional transitions  
of asymptotic particle states whose intrinsic thermodynamics is of no use to describe such processes.
Yet, as we will argue in the present work, defining particle properties can be obtained from a thermodynamical approach operating in their restframes.

A situation that QED and the SM is unlikely to describe well, however, is the physics of a dense and thermal\footnote{Here, {\sl dense} refers to densities of the order of 1/(droplet volume), namely $\sim 1.8\times 10^{28}\,$m$^{-3}$, and {\sl thermal} corresponds to 
temperatures comparable to the stabilisation temperature of the associated Yang-Mills theories, namely $T_0 \sim 9.4\,$keV, or an energy density $\sim 5 \times 10^{14}\,$J m$^{-3}$.}
state of electrons and positrons in terms of particle and energy transport, charge and spin fluctuations as well as  correlation functions of particle number densities and particle speeds. 
An important signature to search for in this regard is the creation of electron-positron pairs by macroscopic droplet evaporation, in terms of their gamma-ray annihilation lines originating outside the droplet. 
The formation of such extended droplets is facilitated by localised deposition of ultra-high energy density through compact, ultra-high contrast, femtosecond lasers focused to relativistic
intensities onto targets composed of aligned nanowire arrays \citep{doi:10.1126/sciadv.1601558}. 
This experiment achieves a local energy density of $8\times 10^{16}\,$J/m$^3$ across 
a spot size of $\sim 5\,\mu$m which would guarantee the creation of macroscopic droplet dimensions.}\\

In this paper, we consider the simultaneous presence of all phases (deconfining\,/\,preconfining\,/\,confining) of 4D SU(2) Yang-Mills thermodynamics within a ball-like spatial region of radius $r_0$, a droplet. In particular, we demonstrate how the mixing within the deconfining phase of two SU(2) Yang-Mills theories of vastly disparate scales, SU(2)$_{\rm CMB}$ and SU(2)$_{\rm e}$, appears to yield a realistic value for the electroweak mixing angle due to vanishing bulk pressure and at vanishing external energy-momentum transfer. The Yang-Mills scales (or critical temperatures) of SU(2)$_{\rm CMB}$ and SU(2)$_{\rm e}$ are derived from experiment. 
Namely, the Yang-Mills scale of SU(2)$_{\rm CMB}$ is suggested to 
be $\Lambda_{\rm CMB}\sim 10^{-4}\,$eV from 
the excess of CMB line temperatures for frequencies $\nu<3\,$GHz \citep{Fixsen:2009xn}
or from a potential detection of screening effects in a laboratory blackbody experiment at low temperatures. 
{ Here, screening effects generate a gap for propagating  blackbody radiation at low frequencies (<\,17\,GHz) shortly above the deconfining/preconfining phase transition of SU(2)$_{\rm CMB}$ \citep{Hofmann:2022qsa,Meinert2024} }
\footnote{This gap is filled by 
evanescent modes with a spectrum differing distinctly from Rayleigh-Jeans: Rather than decaying $\propto\nu^2$ with decreasing frequency $\nu$ the evanescent mode spectrum grows because modes with decreasing $\nu$ possess a decreasing 
spatial decay length, therefore less energy, and thus are easier excitable.}. 

The Yang-Mills scale of SU(2)$_{\rm e}$ is deduced in the present work from the mass of the electron as $\Lambda_{\rm e}\sim 3.60\,$keV, see also \citep{Hofmann:2017lmu,ThierryHofmann2022} for mildly deviating earlier estimates that neglected gauge-theory mixing.
The mixing between the deconfining, preconfining, and confining phases of SU(2)$_{\rm e}$ and the deconfining phase of SU(2)$_{\rm CMB}$ creates a thick boundary shell to the droplet whereas the bulk region is described by the deconfining phases of SU(2)$_{\rm e}$ and SU(2)$_{\rm CMB}$ trapping an SU(2)$_{\rm e}$ monopole\footnote{This requires an electric-magnetic dual interpretation of U(1) charges in SU(2) Yang-Mills theory, see \citep{Hofmann:2012xp,bookHofmann}.}. The thick boundary shell thus is a region of high electric conductivity due to condensed monopoles in the preconfining phase of SU(2)$_{\rm e}$. 
It is an essential assumption of the present paper that in the limit of a static equlibirium the thick boundary shell is sharply delineated from the bulk at $\bar{r}<r_0$. That is, for each volume element within the thick boundary shell contracting forces, stemming from the mixing of deconfining and preconfining phases of SU(2)$_{\rm e}$, are cancelled  by the expanding force due to the plasma of the deconfining phase of SU(2)$_{\rm CMB}$\footnote{To avoid making this assumption a treatment of the thick boundary shell in terms of imperfect-fluid hydrodynamics would require a derivation of an effective, selfconsistent, and $r$ dependent equation of state by a spatio-temporal coarse-graining of the phase and gauge-theory mixture. On this basis, the solution to the hydrostatic fluid equations of the thick boundary shell could be obtained. This is technically hard and beyond the scope of the present paper, however.}. In any case, the pressure vanishes on the confining-phase side of the Hagedorn transition in SU(2)$_{\rm e}$, that is, outside the droplet for $r>r_0$ \citep{bookHofmann}.\\

We also assume (i) effective hydrostatic equilibrium of the thick boundary shell. After spatio-temporally averaging, the bulk pressure $P_{\rm bs}$ vanishes, $P_{\rm bs}\equiv 0$, 

(ii) that phase segregation and therefore a maximum radius for radial averages over monopole positions is given by the mean radius $\bar{r}$ of the initial, 
non-phase segregated thin-boundary shell droplet, and 

(iii) that the energy density within the thick boundary shell is comparable to the bulk energy density. 

Notice that in the absence of an external force bulk homogeneity $T(r)=T_0=\mbox{const}$ and therefore vanishing bulk pressure, $P_{\rm bulk}\equiv 0$, are due to hydrostatic and thermodynamic equilibrium and the continuity condition of vanishing pressure at bulk-boundary-shell separation $r=\bar{r}$, $P_{\rm bulk}$(r=$\bar{r}$)=$P_{\rm bs}(r=\bar{r})=0$. Namely, $dP_{\rm bulk}/dr=dP_{\rm bulk}/dT dT/dr\equiv 0$ $\Rightarrow$ $dT/dr\equiv 0$ $\Rightarrow$ $T(r)=T_0=\mbox{const}$ which, together with $P_{\rm bulk}(r=\bar{r})=P_{\rm bs}(r=\bar{r})=0$, yields $P_{\rm bulk}\equiv 0$.  

{ On one hand, the quantum mass of the droplet is $m_0=\omega_0$, monopole breathing associating with the (circular) frequency $\omega_0$ \cite{DeBroglieRev, Hofmann:2017lmu, ThierryHofmann2022}. On the other hand, the spatial extent of the electron refers to a dynamical equilibrium which is very close to the static 
equilibrium assumed in (i). This can be pictured as follows: The thick boundary shell could effectively produce negative pressure by an imbalance of the partial pressures stemming from phase mixing in SU(2)$_{\rm e}$ (negative) and gauge theory 
mixing with SU(2)$_{\rm CMB}$ (positive). 
The bulk surface would then have to impose a positive counter 
pressure (homogeneous in the bulk due to the quantum correlation length $\mid\phi\mid^{-1}$ being much larger than the droplet's extent), that is $T>T_0$. Alternatively, the thick boundary shell could exhibit an imbalance of partial pressures to effectively produce positive 
pressure to which the bulk surface would react by pull (negative pressure). 
This describes the droplet's breathing around a stable equilibrium where bulk and boundary-shell pressures independently vanish.
In the absence of external forces, breathing amplitudes for the deviations of droplet radii and pressures away from this equilibrium are small\footnote{It is the monopole with a mass fraction of less than 
10\,\% \citep{Hofmann:2017lmu} that drives breathing and, due the defined phase structure of SU(2)$_{\rm e}$, the boundary shell cannot compress/expand itself indefinitely.}. 
Therefore, assuming an overall vanishing of pressures is justified for an initial estimate of monopole charge distribution, bulk gauge-theory mixing, and droplet mass.
Regarding assumption (ii), we argue in Sec.\,\ref{BCTH} that shortly after pair creation in each droplet Yang-Mills phases are not segregated and that therefore, the only radial scale $\bar{r}$ to determine 
segregation is the radial mean subject to a homogeneous probability density over the droplet volume. Assumption (iii) allows for a quick estimate of droplet mass, see Eq.\,(\ref{massfromm0mixing}), since for the boundary shell it is a priori not clear what the effective mixing angle between SU(2)$_{\rm CMB}$ and SU(2)$_{\rm e}$ and what the effective phase mixing within SU(2)$_{\rm e}$ are. 
Note that the fraction of boundary-shell volume to droplet volume is $\sim 27/64\sim 0.42$ and that therefore deviations of the true energy density within the boundary shell from the assumed bulk energy density should not influence the droplet mass estimate on the right-hand side of Eq.\,(\ref{massfromm0mixing}) strongly.}\\ 

It is the equivalent descriptions of the system in terms of effective (and only implicit quantum) bulk thermodynamics on one hand and explicit quantum physics on the other hand that allow to address essentials of the droplet physics, see Eq.\,(\ref{massfromm0mixing}). 
The purpose of the present paper is to demonstrate in such a context what assumptions (i), (ii), and (iii) actually imply for the values of the dimensionless parameters in electroweak theory. 

This paper is organised as follows. For the benefit of readability and better access to our derivations we review in Sec.\,2 the thermal phase structure of a (3+1)-dimensional Minkowskian and effectively quantum SU(2) Yang-Mills theory as it emerges from a 4-dimensional, fundamental, and classical SU(2) Yang-Mills theory defined on the Euclidean cylinder ${\bf R}^3\times S_1$. In particular, we discuss the nature of the transitions from deconfining via preconfining to the confining phase and how the respective (thermal) ground states emerge by dense packings of fundamental topological defects whose central regions are not resolved in the respective effective theories. Sec.\,3 is devoted to a brief review of previous work on finite-volume SU(2) Yang-Mills thermodynamics inside the droplet that represents the spatial region of selfintersection of a center-vortex loop, immersed into the confining phase.  Such a droplet traps a perturbed Bogmolnyi-Prasad-Sommerfield (BPS) monopole causing monopole and plasma vibrations. A stepwise approach towards a droplet model capable of approximating the experimental value of the electromagnetic fine-structure constant $\alpha$ is developed in Sec.\,4. This model, which is short of mixing two SU(2) gauge theories within the droplet's bulk, invokes all three phases of SU(2) Yang-Mills thermodynamics. It also uses a spherical mirror-charge construction to yield the effective droplet charge as seen by an external electromagnetic probe of long wavelength. Moreover, the model introduces the mean of the radial position of the monopole over the entire droplet volume to segregate a thick boundary shell from deconfining-phase bulk thermodynamics. Finally, the droplet charge is expressed as a bulk-volume average.  
In Sec.\,5 the mixing of two SU(2) Yang-Mills theories is considered to describe the thermodynamically  stable bulk of a droplet. Here, the conditions of monopole-charge universality and zero bulk pressure determine both the bulk temperature in units of one theory's critical temperature for the deconfining-preconfining phase transition as well as the mixing angle. The latter turns out to be close to experimental values of the weak mixing angle $\theta_{\rm W}$, the Weinberg angle. When accounting for gauge-theory mixing, the model developed in Sec.\,4 yields a value of $\alpha$ not far from the experimental low-energy value. Sec.\,5 computes the ratio between reduced Compton radius $r_c$, which is roughly equal to the core radius of the monopole, and droplet radius $r_0$ from a quantum-thermodynamical electron-mass formula. As it turns out, $r_0$ is smaller but comparable to the Bohr radius $a_0$. In Sec.\,6 we present a brief summary and an outlook on how the present framework can be applied to understand the emergence and weak decay of the other two charged lepton species of the SM. 

\section{Three phases of SU(2) Yang-Mills thermodynamics\label{3pYMTD}}

\begin{figure}[ht]
\centering
\includegraphics[width=\columnwidth]{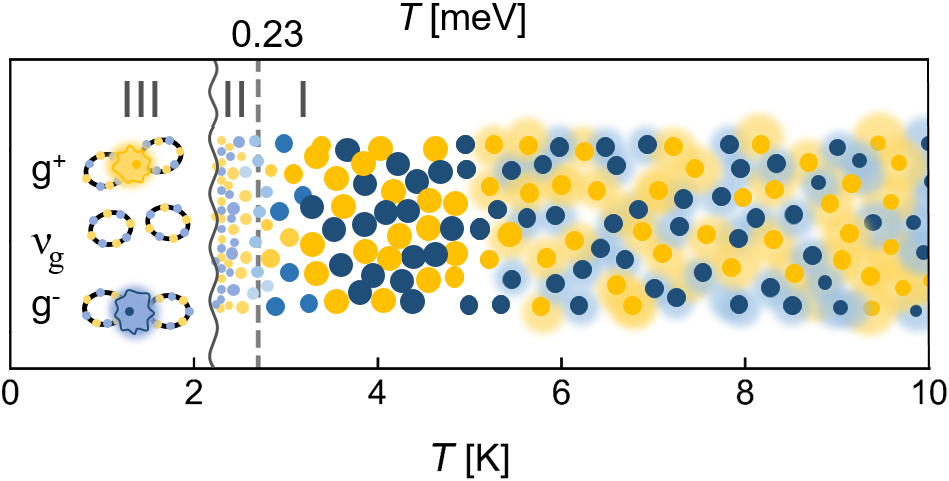}
\caption{Phase diagram of SU(2)$_{\rm CMB}$ with $T_c=2.725\,$K in the infinite-volume limit. There are three distinct phases: (I) deconfining phase, (II) preconfining, and (III) confining phase. In phase III a very light lepton family emerges whose charged members can be called gammaron and antigammaron and which posses a mass of 1.5\,$\times\,10^{-2}$\,eV. The ground state of SU(2)$_{\rm CMB}$ partly transitions into such charged leptons at the Hagedorn temperature $T_{c^\prime}\sim 2.27\,$K.}
\label{Fig:PhaseDiagram}
\end{figure}

SU(2) Yang-Mills thermodynamics occurs in three distinct phases as shown in Fig.\,\ref{Fig:PhaseDiagram}. Here, we briefly review associated concepts and results, closely following \cite{bookHofmann}. The infinite-volume limit is assumed throughout. 

\subsection{Deconfining phase (I)}

The deconfining phase takes place for $T\ge T_c$. Its thermal ground state can be derived from a spatially coarse-grained two-point field-strength correlator, evaluated on a trivial-holonomy caloron or anticaloron \citep{Harrington:1978ve} whose respective contributions are superimposed. The reason for the use of a trivial-holonomy rather than a nontrivial-holonomy caloron or anticaloron is its stability under one-loop quantum fluctuations \cite{Diakonov2004}. (Anti)calorons are (anti)selfdual solutions to the fundamental Yang-Mills equations on the Euclidean cylinder ${\bf R}_3\times S_1$, parametrised by $0\le r<\infty, 0\le\varphi\le 2\pi, 0\le\theta\le\pi, 0\le\tau\le\beta\equiv\frac{1}{T}$ ($T$ denoting temperature). Therefore, these gauge-field configurations are void of pressure and energy-density, that is, they do not propagate. The spatial coarse-graining is performed over the central spatial dependence of the field-strength correlation. This amounts to an integral over the 3D spatial ball, referred to as 'center' in the following, at any given value of $\tau$. In the singular gauge that the (anti)caloron is constructed in, this ball centrally locates the topological charge. Specifically, this means that integrating the Chern-Simons current over a 3-sphere of radius $\rho\gg\epsilon>0$, centered at the position of the maximum of the action density at $r=0,\tau_0$, yields a result which does not depend on $\epsilon$. Here $\rho$ denotes the (anti)caloron radius\footnote{Throughout the paper the symbol $\rho$ is used to denote various quantities in different contexts: caloron radius, plasma energy density, and probability density for the location of a monopole within the droplet's bulk. What is meant when will be clear from the context.}. The average over (anti)caloron radii $\rho$ yields a normalisation which cubically rises with the upper cutoff. This integral produces a rapidly saturating, harmonic $\tau$-dependence. 

Dense spatial packing of centers (spatial homogeneity at the resolution set by a (anti)caloron center's radius) gives rise to an inert, temporally winding, and adjoint scalar field $\phi$ of 
modulus $|\phi|=\sqrt{{\Lambda^3}/{2\pi T}}$ which breaks the fundamental gauge symmetry SU(2) down to U(1). Here $\Lambda$ denotes the Yang-Mills scale of the deconfining phase. Dense packing implies the overlap of (anti)caloron peripheries (the complements of their centers) with centers and with one another. After spatial coarse-graining, the collective presence of all other (anti)caloron peripheries 
at the position of a given (anti)caloron center accurately is captured by a pure-gauge solution $a^{\rm gs}_\mu$ 
to the effective Yang-Mills equation. Out of vanishing pressure and energy density in a situation, where only the dense packing of centers is considered, the inclusion of peripheral overlap produces finite ground-state energy density $\rho^{\rm gs}$ and pressure $P^{\rm gs}$. One has 
\eqb
\label{enedPgs}
\rho^{\rm gs}=-P^{\rm gs}=4\pi\Lambda^3 T\,.
\eqe 
This vacuum equation of state is an important aspect of the `thermal ground state'. Microscopically, (anti)caloron overlap transiently changes the (anti)caloron's holonomy from trivial to mildly nontrivial which introduces the negative ground-state pressure described by Eq.\,(\ref{enedPgs})\cite{Diakonov2004}.  

By virtue of the adjoint Higgs mechanism invoked by field $\phi$, two out of three components (dimension of SU(2) algebra $\mathfrak{su}$(2) equals three) 
of the effectively propagating gauge field $a_\mu$ are massive where 
mass $m$ depends on $T$ (quasiparticle mass). This can be seen after de-winding the field $\phi$ and nullifying the effective ground-state gauge field $a^{\rm gs}_\mu$ by a singular but admissible gauge transformation. Under such a gauge transformation, the Polyakov loop, evaluated on $a^{\rm gs}_\mu$, changes its value from the center element $-{\bf 1}_2$ to the center element ${\bf 1}_2$. This confirms that the theory is in a deconfining phase, that is, the electric center symmetry ${\bf Z}_2$ is broken dynamically.  

Effective, propagating excitations can be grouped into purely quantum thermal for all frequencies 
(massive modes) and classical, off-shell or quantum thermal, depending on frequency (massless mode). The one-loop approximation to thermodynamical bulk quantities like 
pressure $P$ or energy density $\rho$, subjecting noninteracting excitations to the presence of the thermal ground state, is 99\,\% accurate. Small corrections to this estimate can be computed in terms of higher loops. They collectively describe 
the effects of rare dissociations by large holonomy shifts in (anti)calorons. The dissociation products are screened monopole-antimonopole pairs. 

\begin{figure}[H]
\centering
\includegraphics[width=\columnwidth]{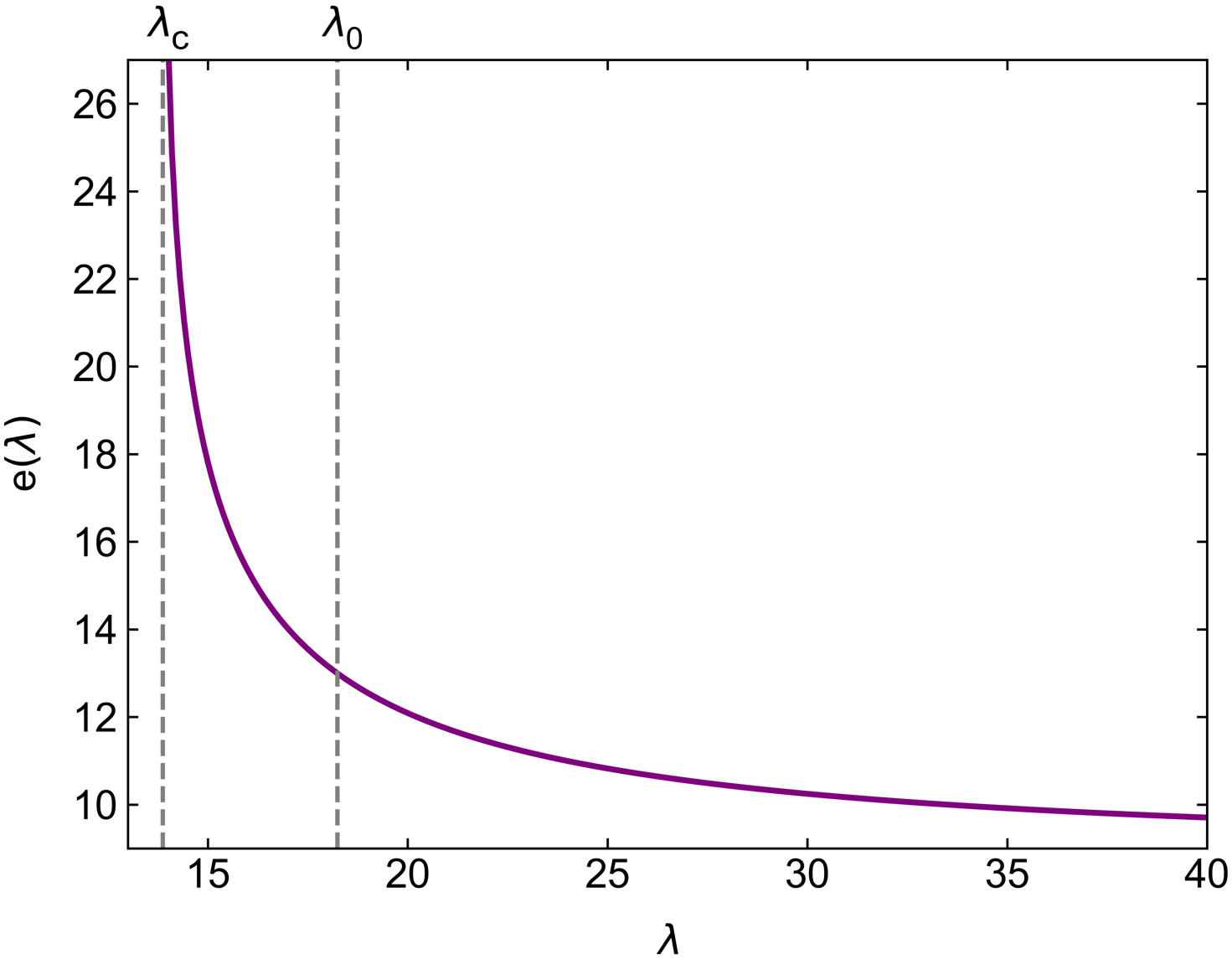}
\caption{The effective gauge coupling of the deconfining phase in SU(2) Yang-Mills thermodynamics as a function of dimensionless 
temperature $\lambda=\frac{2\pi T}{\Lambda}\ge 13.87=\lambda_c$ where $\Lambda$ denotes the Yang-Mills scale, 
and $T_c$ is the critical 
temperature for the deconfining-preconfining phase transition. Temperature $\lambda_0=18.31$ refers to the point of vanishing pressure.}
\label{Fig:Coupling}
\end{figure}

Demanding that one-loop fluctuations and the thermal ground-state estimate are 
thermodynamically consistent (implicit $T$-dependences cancel in Legendre transformations: 
$\frac{dP}{dm}=0$, $P$ the one-loop pressure, $m$ the quasiparticle mass) one derives the following 
first-order ordinary differential equation \citep{bookHofmann}
\begin{equation}
\label{evoleqsu2}
\partial_a\lambda=-\frac{24\lambda^4
  a}{(2\pi)^6}\frac{D(2a)}{1+\frac{24\lambda^3a^2}{(2\pi)^6}D(2a)}\,,
\end{equation}
where 
\eqb
\label{ae}
a\equiv\frac{m}{2T}=2\pi e\lambda^{-3/2}
\eqe
and 
\eqb
D(y)\equiv\int_0^\infty
dx\,\frac{x^2}{\sqrt{x^2+y^2}}\frac{1}{\exp\left(\sqrt{x^2+y^2}\right)-1}\,.
\eqe
Here, $e$ denotes the effective gauge coupling in the 
deconfining phase, and dimensionless temperature is defined as 
$\lambda\equiv\frac{2\pi T}{\Lambda}$. The
evolution described by Eq.\,(\ref{evoleqsu2}) possesses two 
fixed points: $a=0$ and $a=\infty$. The latter associates with 
a critical temperature $\lambda_c$ of value $\lambda_c=13.87$, the former describes the behaviour at asymptotically high temperatures.  
For $a\ll 1$ the downward-in-temperature evolution described by Eq.\,(\ref{evoleqsu2}) is solved by
\begin{equation}
\label{solasu2} 
a(\lambda)=4\sqrt{2}\pi^2\lambda^{-3/2}\left(1-\frac{\lambda}{\lambda_i}\left[1-\frac{a_i\lambda_i^3}{32\pi^4}\right]\right)^{1/2}\,, 
\end{equation}
subject to the initial condition $a(\lambda_i)=a_i\ll 1$ and $\lambda_i\gg 1$. 
The attractor $a(\lambda)=4\sqrt{2}\pi^2\lambda^{-3/2}$ implies, by virtue of Eq.\,(\ref{ae}), that the effective gauge coupling $e$ is a constant almost everywhere: 
\eqb
\label{plateau}
e\equiv\sqrt{8}\pi\,.
\eqe
Eq.(\ref{solasu2}) indicates that the condition $a\ll 1$, under which it was derived, is violated at small temperatures. Since the exact solution continues to grow with decreasing $\lambda$ (negative definiteness of
right-hand side of Eq.\,(\ref{evoleqsu2})) the right-hand side of 
Eq.\,(\ref{evoleqsu2}) will be suppressed exponentially. This implies 
a behaviour $a(\lambda)\propto -\log(\lambda-\lambda_c)$ for
$\lambda\searrow\lambda_c$ and therefore a logarithmic singularity at $\lambda_c$ also for
$e(\lambda)$, see Fig.\,\ref{Fig:Coupling}. Physically, such a 
singularity implies masslessness for isolated screened monopoles, liberated by the dissociation of large-holonomy (anti)calorons and collectively associable to effective radiative corrections \citep{Ludescher:2008cq}. Therefore, as $\lambda\searrow\lambda_c$, the holonomy of a typical (anti)caloron moves from small  
to large \citep{Gerhold_2007}.

\subsection{Preconfining phase (II)}

The preconfining phase extends from $\lambda_c=13.87$ (onset of monopole condensation, second-order phase transition, pressures on both side of the phase boundary match, thermal energy density of monopole-antimonopole-condensed system with one massive 
photonic vector excitation is higher on the preconfining than on the deconfining side of the phase boundary with one massless photonic vector excitation, coexistence of deconfining and preconfining phases, phase tunneling) via $\lambda=\lambda_*=12.15$ (preconfining-phase energy density matches supercooled thermal energy density of deconfining ground state with one massless photonic excitation, phase tunneling suppressed) to $\lambda_{c^\prime}=11.57$ (massive photon decouples, entropy vanishes, onset of Hagedorn transition, discontinuous phase changes of order parameter \cite{tHooft:1977nqb} associate with tunneling towards (magnetic) ${\bf Z}_2$ degenerate ground state, associated center-flux creation). 
Microscopically, the thermal ground state of the preconfining phase is constituted from  massless magnetic monopoles and antimonopoles.
In physics models, these are dually interpreted and therefore represent electric monopoles  and antimonopoles \citep{Hofmann:2012xp}.

Dense packing of monopole and antimonopole cores is described by an effective, inert, complex scalar field $\varphi$ which breaks the remaining gauge symmetry U(1) dynamically.
The overlap of all peripheries at the position of a given monopole or antimonopole core is described by an effective pure-gauge configuration $a_\mu^{D,{\rm gs}}$. 
In the preconfining phase, the Polyakov loop, evaluated on the effective ground-state gauge field 
$a_\mu^{D,{\rm gs}}$, is unity in both winding and unitary gauge. 
This is indicative of the fact that the preconfining phase already confines infinitely heavy, fundamental test charges, although massive, propagating gauge modes can still be excited.  

The pressure throughout the deconfining phase is {\sl negative}. If it were not for phase mixing, the ground state of the preconfining phase would be a superconductor. But even in the presence 
of phase mixing, electric conductance is expected to be very high for 
$\lambda_{c^\prime}\le \lambda\le \lambda_c$ compared to the deconfining and confining phases. 
At $\lambda_{c}$ the effective U(1) gauge coupling $g$ vanishes with mean-field exponent $\frac12$, at $\lambda_{c^\prime}$ $g$ diverges logarithmically \citep{bookHofmann}.

\subsection{Confining phase (III)}

At $\lambda_{c^\prime}$ massive Abrikosov-Nielsen-Oleson 
vortex loops without selfintersections, which, due to their finite, resolvable core-sizes, are instable defects for $\lambda_c^{\prime}<\lambda\le\lambda_c$, become massless and metastable. These massless solitons, so-called thin center-vortex loops, break the magnetic center symmetry ${\bf Z}_2$ dynamically \cite{tHooft:1977nqb}  
upon their condensation. Condensation is enabled by shrinkage to round, massless points \cite{Moosmann:2008fr}. The ensuing new ${\bf Z}_2$ degenerate ground state, 
which confines fundamental test charges and does not support thermal gauge-mode excitation anymore, can be shown to possess no energy density and to exert no pressure \citep{bookHofmann}. 

The liberation of thin center-vortex loops at $\lambda_{c^\prime}$ from the ground state of the preconfining phase initiates the Hagedorn transition. Effectively, this process is described in terms of phase changes by $\pm\pi$ of a complex scalar field $\Phi$.Such phase changes accompany tunneling events through tangentially convex regions of a potential uniquely determined by ${\bf Z}_2$ symmetry. Collisions of center-vortex loops lead to twistings and the formations  
of stable regions of selfintersections which, in their restframes, are characterised by the droplets of radius $r_0$ that we have alluded to in Sec.\,\,1, see also Sec.\,3. The density of states $\Omega(E)$ of solitons, which are subject to an arbitrary number of selfintersections, rises more than exponentially in energy $E$. Therefore, no partition function exists for $\lambda\sim\lambda_{c^\prime}$ (confining phase), and thermodynamics is not a valid description anymore. (This can already be inferred from the fact that for $\lambda\searrow \lambda_{c^\prime}$ the entropy density vanishes which violates Nernst's theorem.)

\subsection{Summary of phase structure of SU(2) Yang-Mills thermodynamics}

To summarise, SU(2) Yang-Mills thermodynamics comes in three phases, see Fig.\,\ref{Fig:PhaseDiagram}. The deconfining phase (phase I) breaks SU(2) to U(1) in terms of a thermal ground state estimate which is composed of trivial-holonomy anticalorons and calorons of topological charge $k=\mp 1$. For $\lambda\gg\lambda_c$ rare dissociations of (anti)calorons, induced by large holonomy shifts, create isolated monopoles and antimonopoles which, however, are spatially not resolved in the effective theory. For $\lambda\sim 2\dots3\,\lambda_c$ the distance between monopoles and antimonopoles is comparable to the spatial resolution set by $|\phi|^{-1}$, and therefore they become explicit and isolated degrees of freedom. In the effective theory, the collective imprint of these monopoles and 
antimonopoles onto thermodynamical quantities (a $\sim 1\%$ effect compared to the free quasiparticle approximation \citep{bookHofmann}) or dispersion laws for propagating gauge modes can be studied in terms of (resummed) radiative corrections. There is a temperature, $\lambda_0\sim 1.32\,\lambda_c$, where a finite-size system can be stabilised due to vanishing pressure. \\  

At $\lambda_c$ monopoles and antimonopoles become massless, pointlike defects that are densely packed: they represent a would-be superconductive condensate. However, phase mixing reduces conductivity to finite values. Pressure is negative throughout the preconfining phase.\\ 

At $\lambda_{c^\prime}=0.83\,\lambda_c$ the monopole-antimonopole condensate of the preconfining phase undergoes a violent decay into center-vortex loops of any selfintersection number $n\ge 0$. This initiates the (nonthermal) Hagdorn transition. The new ground state is composed of massless, shrunk-to-points center-vortex loops, possesses no energy density and exerts no pressure. Each selfintersection point of a center-vortex loop represents the droplet of radius $r_0$ (scaled by the ratio of the respective Yang-Mills scales) that is addressed in the next section. At a finite resolution, provided  
by nonthermal, external electromagnetic fields, only center-vortex loops with $n=0,1$ can be regarded stable solitons.

\section{Selfintersecting center-vortex loop with 
\texorpdfstring{$n=1$}{n=1} and magnetic moment of quantum soliton
\label{sicvl}}

A thin center vortex can be understood as a chain of unresolved monopoles and antimonopoles whose flux is confined to a thin tube \citep{DelDebbio:1997ke}. The stable, massive soliton, represented by a 
center-vortex loop with $n=1$, then originates from a localised investment 
of energy (pair creation) into round-point center-vortex loops with $n=0$. These round points  constitute the ground state of the confining phase. Such an investment of energy implies stretching / twisting / pinching and the eventual release of a monopole or an antimonopole from the tube. As a consequence, points of vortex selfintersection are defined. Since a monopole or antimonopole cannot exist as isolated defects in the confining vacuum, each point of selfintersection needs to evolve into an extended ball-like spatial region (a droplet).  The deconfining, pressureless bulk of such a droplet facilitates a finite-mass and finite-extent monopole or antimonmopole at the temperature $T=T_0=1.32\,T_c$. The droplet comprises a boundary shell, which exhibits a temperature gradient, where positive pressure generated by turbulences (Hagedorn transition) is superimposed by the negative pressures contributed by preconfining and deconfining phases. On temporal and spatial average, zero pressure of the boundary shell is assumed in the remainder of this paper to continue the vanishing pressure of the confining phase outside the droplet. The center flux external to the droplet forms a flux configuration which is of figure-eight topology. \\

\begin{figure}[ht]
\centering
\includegraphics[width=\columnwidth]{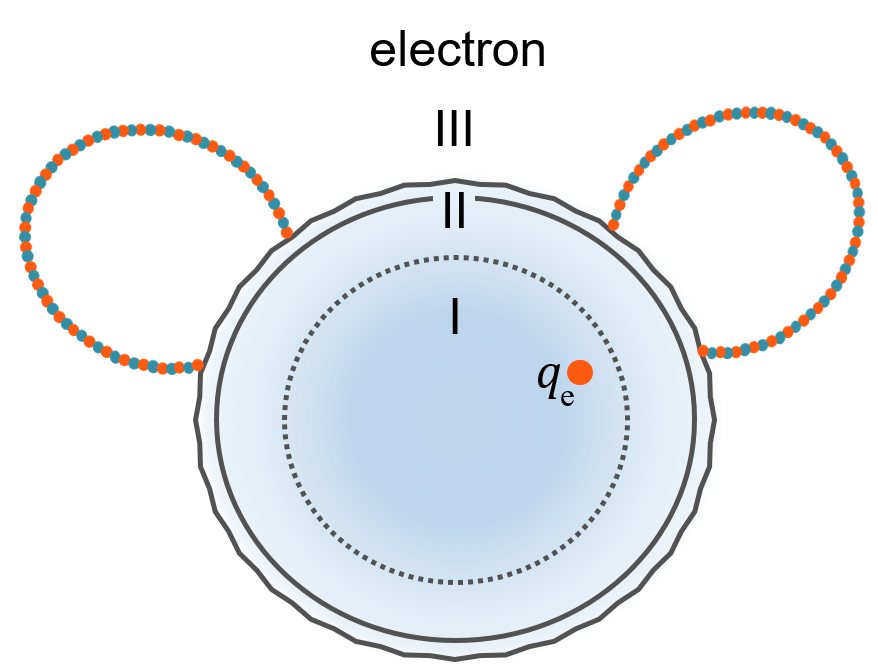}
\caption{The electron as a soliton in SU(2) Yang-Mills theory. For simplicity, mixing effects with another SU(2) Yang-Mills theory are ignored in this figure. The bulk of the droplet, containing an isolated BPS monopole, is in the deconfining phase (I) at vanishing pressure (temperature $T_0$), the thick boundary shell (II) radially interpolates between a thermal 2nd-order like (inner part of shell, onset of superconductivity / preconfining phase) and a nonthermal Hagedorn phase transition (outer part of shell, full superconductivity). On average, this boundary shell should be of vanishing pressure due to the mixing of all three phases and possesses a high electric conductance. The confining phase (III) prevails outside of the droplet. Isolated gauge modes may propagate as waves in this phase if their frequency is lower than a limit set by the Yang-Mills scale. The monopole within the bulk of the droplet breathes at a frequency $m_{\rm e}$ (rest mass of electron \citep{Hofmann:2017lmu}) and generates a time dependent vibration of spatially constant amplitude throughout the droplet. Connected to the thick boundary shell are center-vortex lines, which induce the magnetic moment of the soliton. Topologically, these form a figure eight with the vortex-line selfintersection - producing an isolated monopole or antimonopole - being responsible for the origination of the droplet, see text.}
\label{Fig:PhaseDiagram2}
\end{figure}

Let us now discuss why such a topology relates to the spin-$\frac12$ nature of the electron and how an according magnetic moment $\vec{\mu}$ emerges. Dually interpreted, a quantum of magnetic center flux represents a quantum of electric center flux, in turn, inducing a quantum of magnetic moment. Fig.\,\ref{Fig:PhaseDiagram2} depicts the center-vortex loop with $n=1$. Since the electric center flux is two-fold degenerate (it can flow along or counter a fixed tangential vector to the vortex loop) the projection of $\vec{\mu}$ onto a quantisation axis is also twofold degenerate, and one has in the isolated case
\eqb
 \label{magmom}
\vec{\mu}=-g\,\mu_{\text{B}}\,\vec{S}\,,
\eqe
where $g=2$, $\mu _{\text{B}}\equiv -\frac{q}{2m_{\rm e}}$ is the Bohr magneton, $m_{\rm e}$ and $q$ are the electron's mass and charge, respectively, and $\vec{S}$ is the spin vector. The two possible projections of $\vec{S}$ are $\pm\frac12$. Notice that the quantisation axis is parallel to the normal $\hat{n}$ of the plane that the figure-eight configuration is considered to be immersed in, for certain characteristics of curve-shrinking planar figure-eight configurations, see \citep{Grayson1989}. 

Picturing the droplet charge, modulo quantum jitter \citep{Breit1928,Schrodinger:1930mpi}, to move along a circle of radius comparable to $a_0$ (Bohr radius), 
see Sec.\,\ref{massformula}, the Bohr magneton describes its effective revolution in time $\Delta t$ as triggered by the revolution within the same span of time of one of the unresolved monopoles or antimonopoles along the thin vortex loop. Namely, in associating with $\vec{L}=-g\vec{S}$ the orbital angular momentum of the droplet subject to $|\vec{L}|=a_0\cdot m_{\rm e}\cdot  2\pi a_0/(\Delta t)$, the magnetic moment $\vec{\mu}$ of Eq.\,(\ref{magmom}) can simply be interpreted as $\vec{\mu}=\pm I\,\pi a_0^2\,\hat{n}$ which is the magnetic moment induced by a circular current loop of radius $a_0$ carrying the current $I=\frac{q}{\Delta t}$. Since a center-vortex loop in reality is not isolated (it is immersed into the CMB, and it connects to a fluctuating thick boundary shell) one expects a slight deviation 
from the value $g=2$ which to a high precision is computable in Quantum Electrodynamics \citep{PhysRev.73.416}. A derivation of $\mu_{\text{B}}$ from jittery revolutions of the droplet is beyond the scope of the present paper, however. 

Ignoring for the moment the complications of electroweak decays, mixing effects, and the tendency of vortex loops to shrink under a lowering of external resolution \citep{Grayson1989}, we propose the three lepton families to match with doublets of center-vortex loops ($n=0,1$). Each doublet would then emerge within one of three SU(2) Yang-Mills theories, whose scales relate to charged lepton masses, 
see Sec.\,5.4 for a derivation of droplet size and $T_c$ from the mass of any given charged lepton. For the first lepton family, where the soliton with $n=1$ (electron) is stable, the implications of such an assignment are pursued in the remainder of this paper. 

\section{Value of electromagnetic fine-structure constant: Effective droplet charge in pure \texorpdfstring{SU(2)$_{\rm e}$}{SU(2)e}\label{chargescreening}}

\subsection{Trapping a single monopole or antimonopole inside the droplet's bulk}

\begin{figure}[ht]
\centering
\includegraphics[width=\columnwidth]{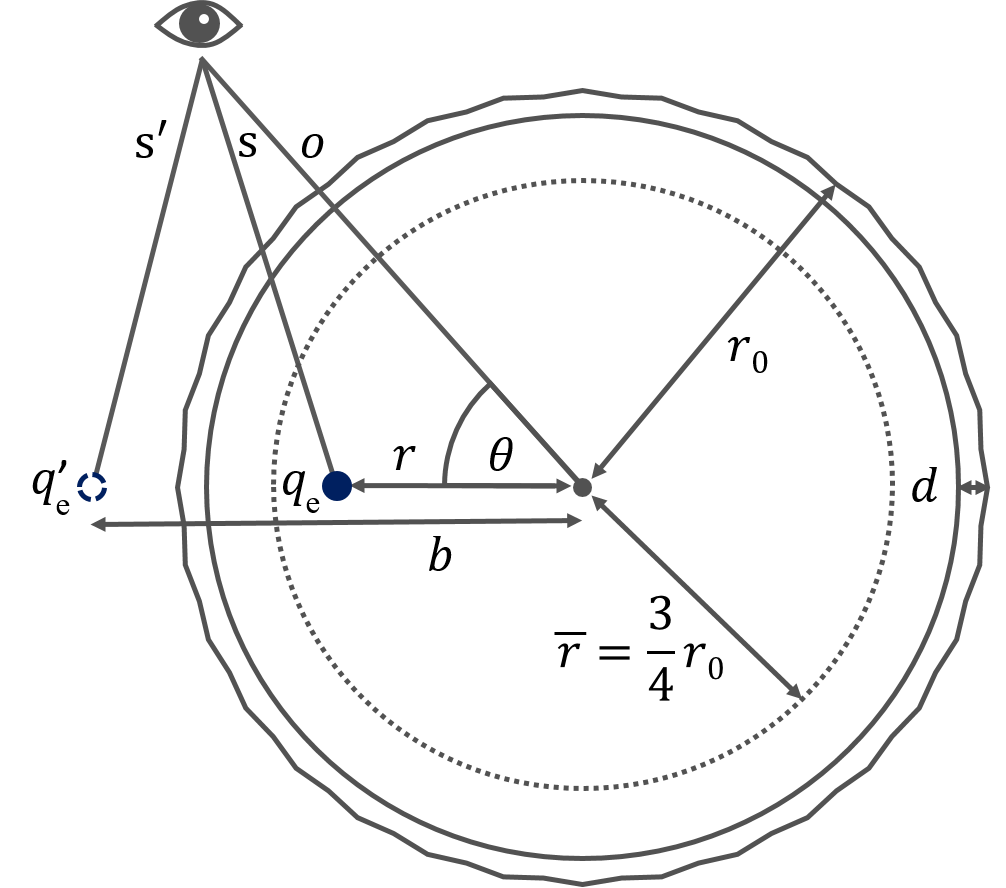}
\caption{Construction of the effective droplet charge by superposition of the Coulomb potentials of the monopole charge $q_{\rm e}$ with a mirror charge $q_{\rm e}^\prime$ outside the droplet due to superconductivity of a thin boundary shell. In reality, a finite radial range $r_0-d\le r \le r_0$ associates with the Hagedorn transition. The reason why $\bar{r}=\frac34\,r_0$ (dotted line) delineates region II from region I, compare with Fig.\,\ref{Fig:PhaseDiagram2}, is discussed in Sec.\,\ref{BCTH}.   
} 
\label{Fig:MirrorCharge}
\end{figure}

Let us first discuss the simplified situation in which the physics of a droplet is governed 
by a single gauge theory: SU(2)$_{\rm e}$. 
We aim to derive the dual charge\footnote{The magnetic charge of the monopole is physically interpreted as an electric charge, see  \citep{Hofmann:2012xp,bookHofmann}.} 
of such a system. A monopole or an antimonopole is immersed into the droplet's bulk of deconfining phase at temperature $T_0$. From now on we consider the situation of a trapped monopole only, since that of a trapped antimonopole simply is described by sign inversions of all charges considered. As discussed previously, $T_0=1.32\,T_{c,\rm e}$ is the temperature where the pressure of the droplet vanishes \cite{ThierryHofmann2022}. Due to the high conductance of the boundary shell ($\bar{r}\le r\le r_0$) and for $\bar{r}\sim r_0$ a spherical mirror-charge construction \citep{MIC} can be used to approximate the droplet charge for probes of long wavelengths. This construction is depicted in Fig.\,\ref{Fig:MirrorCharge}. Here $s$ or $s^\prime$ denote the respective distances between observer and monopole or observer and mirror charge. Monopole and mirror charge are both positioned on a radial ray pointing away from the droplet center. As seen from the droplet center, angle $\theta$ subtends the direction of the observer and the direction of this ray. The distance between observer and droplet center is denoted by $o$. In thin-shell approximation, we set $\bar{r}=r_0$.\\ 

Since a mirror charge construction for boundary conditions on a spherical surface operates  
with Coulomb potentials for the inducing and the mirror charge \citep{MIC}, it is essential to secure that a Yukawa factor $\exp(-s/l_s)$ for the potential associated with the monopole's  charge inside the ball can be treated as unity. 
Here $l_s$ denotes the charge screening length that arises from other screened and stable 
dipoles in the infinite-volume plasma \citep{bookHofmann}. Let us thus check the selfconsistency of only one explicit monopole or antimonopole inhabitating the droplet of radius $r_0$. This requires an estimate of $l_s/r_0\gg 1$. In \citep{ThierryHofmann2022} it was found for a pure SU(2)$_{\rm e}$ theory describing 
the droplet that 
\eqb
\label{r0SU2epure}
r_0=\frac{4.043}{\pi T_0}=0.44\,\Lambda_{\rm e}^{-1}\,. 
\eqe 
In Sec.\,\ref{massformula} we derive a deviating estimate of $r_0$ which includes the mixing of SU(2)$_{\rm CMB}$ and SU(2)$_{\rm e}$. 
The screening length $l_s={71.43\, /\, T}$ was extracted from a two-loop radiative correction 
to the pressure at large temperatures, $T\gg T_0$, see \citep{Ludescher:2009my}. For an estimate, we may continue this asymptotic result to $T_0$ to obtain
\eqb
\label{lsvsr0}
\frac{l_s}{r_0}\sim 55.5\gg 1\,.
\eqe 
Therefore, for $0\le s\le 2\,r_0$ a Yukawa factor $\exp(-s/l_s)$ to the Coulomb potential, which could arise from other explicit monopoles and antimonopoles within the droplet, can be set equal to unity. Thus, the assumption of a single monopole being trapped in the bulk of the droplet indeed is selfconsistent. 

\subsection{Thin-shell approximation\label{MCC}}

In deriving the effective charge of the droplet at bulk temperature $T=T_0$ and in the long-wavelength limit, we first analyse the limits of vanishing shell thickness $r_0=\bar{r}$ and superconductivity of the boundary shell, see Fig.\,\ref{Fig:PhaseDiagram2}. Moreover, we neglect the effects of monopole breathing \citep{Fodor_2004,Forg_cs_2004} and implied position changes in this section. Under these simplifying assumptions, the static potential outside the droplet reads \citep{MIC}
\eab
\label{Vr}
V(o,r,\theta)&=&\frac{1}{4\pi}\left(\frac{q_{\rm e}}{s}+\frac{q_{\rm e}^\prime}{s^\prime}\right)\nonumber\\
&=&\frac{q_{\rm e}}{4\pi}\left(\frac{r-r_0}{r}\frac{1}{o}+\frac{(r^2-br_0)\cos\theta}{r}\frac{1}{o^2}\right.\nonumber\\
& &+\,\left.\frac{(r^3-b^2r_0)(1-3\cos^2\theta)}{2r}\frac{1}{o^3}\right.\nonumber\\ 
& &+\,\left.{\cal O}\left(\frac{1}{o^4}\right)\right)\,,\ \ (o>r_0)\,,
\eae
where $s\equiv\sqrt{o^2+r^2-2or\cos\theta}$, $s^\prime\equiv\sqrt{o^2+b^2-2ob\cos\theta}$, $b\equiv\frac{r_0^2}{r}$, and $q_{\rm e}^\prime\equiv -q_{\rm e}\frac{r_0}{r}$.  
From Eq.\,(\ref{Vr}) and for $o\gg r_0$ we read off the effective charge $g_{\rm e}$ of the droplet from the coefficient 
of $\frac{1}{o}$ as
\eqb
\label{effCharge}
g_{\rm e}=\frac{r-r_0}{r}\,q_{\rm e}\,.
\eqe

\subsection{Averaging the droplet's effective charge over monopole locations}

\subsubsection{Droplet creation and thermalisation\label{BCTH}}

When a droplet of radius $r_0$, containing a trapped monopole, forms in an SU(2)$_{\rm e}$ gauge theory it is a priori not thermalised, and the monopole's location 
within the droplet, which is a mixed state of all three phases, is equally likely everywhere in the ball apart from a thin boundary shell. A uniform a priori distribution of monopole location in Cartesian coordinates is then cast into a nonflat probability density 
$\rho(r,\theta)$ in spherical coordinates for $0\le r<r_0$ and for a polar angle $0\le\theta<\pi$, see Sec.\,\ref{ntsa}. The thin boundary shell 
of the droplet (the radial location $r\sim r_0$ where the Hagedorn transition to the confining phase takes place) is characterised by decoupled dual Abelian 
gauge modes and off-Cartan modes which, due to their large masses, cannot be redistributed into the interior of the droplet. Only after bulk thermalisation is attained for a central radial region does an extended boundary region materialise -- a thick boundary shell ($\bar{r}=\xi r_0<r<r_0$ with $0<\xi<1$) exhibiting a temperature gradient and defining the transition from the deconfining via the preconfining to the confining phase via phase mixing. As the trapped monopole enters this region, it rapidly changes its identity\footnote{The monopole is exposed to the caloron's 4D winding number, which is localised spatially deeply within its center \cite{bookHofmann}. This 4D Euclidean winding is associated with the quantum of action $\hbar$ \cite{bookHofmann} and therefore exerts undeterministic kicks and position changes onto the monopole at the breathing frequency of the monopole.}.\\ 

Let us discuss this situation and its consequences in more detail. If, after bulk thermalisation, the monopole is kicked towards the thick boundary shell then it undergoes mass and charge reduction until it becomes a part of the (spatially inhomogeneous) monopole-antimonopole condensate contributing to the phase mixture there \citep{bookHofmann}. 
Yet, to explicitly conserve charge (or 3D winding number) and droplet mass, another monopole, formerly part of the condensate, needs to act in place of the original monopole in the bulk of the droplet (deconfining phase). 
Ignoring correlations between monopoles and antimonopoles inside and close to the condensate within the boundary shell, 
one may approximate a given monopole's probability density of location by the 
a priori distribution $\rho(r,\theta)$. However, the monopole's charge $q_{\rm e}$ (and its mass) is a function of $r$ which is flat and finite in the 
droplet's bulk (deconfining phase) but rapidly approaches zero as the 
radial location approaches $\bar{r}=\xi r_0$. Since no length scale other than $r_0$ characterises 
the a priori distribution $\rho(r,\theta)$ in the limit of vanishing probe four-momentum transfer, we are 
led to associate $\xi$ with the mean radial position of the monopole (fluctuations of this 
mean can be neglected at vanishing four-momentum transfer, see Sec.\,\ref{ntsa}) according to $\rho$, 
\eqb
\label{emergenceofxi}
\frac{\langle r \rangle_\rho}{r_0}\equiv \xi=\frac34\,. 
\eqe
Physically, the droplet's bulk is subject to a mixing of two thermal gauge 
theories, SU(2)$_{\rm e}$ and SU(2)$_{\rm CMB}$, see Sec.\,\ref{WA}. This, however, does 
not invalidate the above argument, since the (nonthermal) probe field always 
resides in SU(2)$_{\rm e}$ \citep{bookHofmann} and therefore is sensitive only to the physics acted out by (thermal) SU(2)$_{\rm e}$ gauge fields inside the droplet.  

\subsubsection{Naive thin-shell approximation and value of fine-structure constant $\alpha$\label{ntsa}}

The droplet's effective charge, when probed with an energy-momentum transfer small compared to the Yang-Mills scale 
$\Lambda_{\rm e}$, due to the frequent changes of monopole location inside the droplet's bulk (kicks issued by a combination  
of Planck's quantum of action $\hbar$ and the 
lowest breathing frequency of the monopole \citep{Fodor_2004,Forg_cs_2004,ThierryHofmann2022,Hofmann:2017lmu}) compared to the frequency of an external probe field, will represent itself as an angular and radial average subject to the following a priori probability density $\rho$ prior to thermalisation (uniform distribution in Cartesian coordinates over ball volume)
\eqb\label{rho}
\rho(r,\theta)=\frac{3r^2}{4\pi r_0^3}\sin{\theta}\,,\ \ \ (0\le r<r_0,\, 0\le\theta<\pi)\,.  
\eqe
Performing the $\theta$-average of $V(o,r,\theta)$ in Eq.\,(\ref{Vr}) w.r.t. the distribution $\rho(r,\theta)$ in Eq.\,(\ref{rho}), 
we conclude that there is no dipole (and no quadrupole) contribution to the electric charge distribution of the droplet -- in agreement with experimental bounds on a potential, feeble CP violation by an electric dipole moment \citep{doi:10.1126/science.adg4084}. 

In particular, the droplet's charge is obtained by an average of the static system's charge in Eq.\,(\ref{effCharge}) w.r.t. the probability density $\rho(r,\theta)$ in Eq.\,(\ref{rho}). It is given as
\eqb
\label{meanvalueMP}
\langle g_{\rm e} \rangle_{\rho}=3\int_0^{r_0} dr\,\frac{r(r-r_0)}{r_0^3}\,q_{\rm e}=-\frac12\,q_{\rm e}\,.
\eqe
Let us discuss why this result is statistically stable. The monopole changes its position within the droplet at a rate comparable to its breathing frequency $\omega_0=m_{\rm e}$ \cite{DeBroglieRev} where $m_{\rm e}$ denotes the electron mass \citep{Fodor_2004,Forg_cs_2004,ThierryHofmann2022,Hofmann:2017lmu}. If the droplet is probed by an external 
electromagnetic wave of frequency $\omega_p\ll\omega_0$ then the distribution $\rho(r,\theta)$ of Eq.\,(\ref{rho}) is sampled independently $N=\frac{\omega_0}{\omega_p}\gg 1$ many times during each probe oscillation. Therefore, the value of $g_{\rm e}$, averaged over one probe oscillation, has mean 
\eqb
\label{chargeav}
\bar{g}_{\rm e}\equiv \frac{1}{N}\sum_{i=1}^{N} \langle g_{e,i}\rangle_{\rho}=\langle g_{e}\rangle_{\rho}=-\frac12\,q_{\rm e}\,,
\eqe 
and standard deviation $\sigma_{g_{\rm e}}/\sqrt{N}$ where $\sigma_{g_{\rm e}}=\frac{\sqrt{3}}{2}\,q_{\rm e}$ is the standard deviation of $g_{e}$ with respect 
to $\rho(r,\theta)$. In the limit of vanishing four-momentum transfer ($N\to\infty$), the droplet's effective charge hence does not fluctuate and is given by Eq.\,(\ref{meanvalueMP}). Such an argument in favour of the non-fluctuativity of the effective chargecan be extended to assure the nonfluctuativity of any observable function $f(r)$ of the random variable $r$. For example, one may consider the mean radial position, $f(r)=r$, see also Eq.\,(\ref{emergenceofxi}), for which one has 
\eqb
\label{radiusav}
\bar{r}\equiv \frac{1}{N}\sum_{i=1}^{N} \langle r_{i}\rangle_{\rho}=\langle r\rangle_{\rho}=\frac34\,r_0\,
\eqe
with a standard deviation $\sigma_{r}/\sqrt{N}$ where $\sigma_{r}=\sqrt{3/80}\,r_0$. 

Since the external gauge field probing the droplet's charge resides in the Cartan algebra of SU(2)$_{\rm e}$ \citep{bookHofmann} and ignoring 
the effects due to a finite boundary-shell thickness, this would yield a value of the inverse electromagnetic fine-structure constant $\alpha^{-1}$ of
\eqb
\label{alphaunmixed}
\alpha^{-1}=\frac{4\pi}{\bar{g}_{\rm e}^2}=4\frac{e^2(\lambda_0)}{4\pi}=53.464\,,
\eqe
where $e(\lambda_0)=12.96$, see \citep{Hofmann:2017lmu,ThierryHofmann2022}. Obviously, this is too low compared to the value $\alpha^{-1}\sim 137$ 
associated with Eq.\,(\ref{definition_alpha}). Therefore, in Sec.\,\ref{thickshellunmixed} we consider a model implementing the effects of a thick boundary shell. 

\subsubsection{Modelling a thick boundary shell after thermalisation\label{thickshellunmixed}}

According to the discussion of Secs.\,\ref{BCTH} and \ref{ntsa} about droplet generation the only length scale  
available to radially separate bulk thermodynamics (deconfining phase) from the 
thick boundary shell, where the transitions from deconfining via preconfining 
to confining phase take place and where an isolated monopole cannot exist, is $\bar{r}=\xi r_0$ with $\xi=\frac34$, see Eqs.\,(\ref{emergenceofxi}) and (\ref{radiusav}). This corrects 
the thin-shell approximation for the droplet charge $q_{\rm e}$ to 
\eqb
\label{meanvalueMPxi}
\bar{g}_{\rm e}=\langle g_{\rm e} \rangle_{\rho}=3\int_0^{\frac34 r_0} dr\,\frac{r(r-r_0)}{r_0^3}\,q_{\rm e}=-\frac{27}{64}\,q_{\rm e}\,.
\eqe
The inverse electromagnetic fine-structure constant $\alpha^{-1}$ would then compute as 
\eqb
\label{alphaunmixed2}
\alpha^{-1}=\frac{4\pi}{{\bar{g}_{\rm e}}^2}
=\left(\frac{64}{27}\right)^2
\frac{e^2(\lambda_0)}{4\pi}=75.099\,.
\eqe
Although this value is higher than the one of Eq.\,(\ref{alphaunmixed}), it is still quite far off the experimental value associated with Eq.\,(\ref{definition_alpha}). One may think that this discrepancy arises because thin-shell perfect-conductor boundary conditions were used to model a thick boundary shell of finite conductance. Certainly, there is a (not easy to compute) correction to Eq.\,(\ref{alphaunmixed2}) to compensate for this simplification. However, a far more important conceptual ingredient is missing in Eq.\,(\ref{alphaunmixed}): Mixing effects due to the two gauge theories SU(2)$_{\rm e}$ and SU(2)$_{\rm CMB}$ providing a stable mix of deconfining-phase plasmas in the bulk of the droplet to yield an environment that allows the existence of an isolated monopole for each theory. It is the SU(2)$_{\rm e}$ component of this mixture whose monopole together with its mirror charge determine the charge distribution of an electron as seen by a directed, external SU(2)$_{\rm e}$ probe field. To analyse this, is the subject of the next section.\\ 

\section{SU(2) gauge group mixing and bulk thermodynamics inside the droplet}

\subsection{Thermal SU(2) gauge-theory mixing and comparison with electroweak theory} 

Outside the droplet (confining phase of SU(2)$_{\rm e}$, deconfining phase of SU(2)$_{\rm CMB}$ due to the presence of the Cosmic Microwave Background), the directedly propagating (nonthermal) gauge field is the effective Cartan gauge field $a^3_{\mu,{\rm e}}$ \citep{Hofmann:2015oia2,bookHofmann}. The Standard Model of 
Particle Physics (SM) describes this nonthermal electromagnetic mode by a global mixture of the gauge mode $B_\mu$ of U(1)$_Y$ and the Cartan gauge mode $W^0_\mu$ of SU(2)$_{\rm W}$. However, due to the finite extent of the deconfining SU(2)$_e$ plasma in the present approach mixing of SU(2)$_{\rm CMB}$ and SU(2)$_{e}$ occurs only inside the droplet due to outside temperatures being much smaller than the critical temperature $T_{c,e}$ of the deconfining-preconfining phase transition of SU(2)$_e$. De-mixing of the two theories outside the droplet takes place because each is in a different phase. Here, we consider a mixing between the full gauge groups SU(2)$_{\rm e}$ and SU(2)$_{\rm CMB}$ inside the droplet. That is, inside the droplet, due to 
thermalisation in one and the same deconfining phase the fields $f_{\rm CMB}$ and $f_{\rm e}$ are actively rotated as
\begin{align}
\label{mixthetaw}
\begin{pmatrix}
           f_{{\rm CMB}}\\
          f_{\rm e}\\
         \end{pmatrix}
         &\longrightarrow\begin{pmatrix}
          \ \ \cos\theta_{\rm W} & -\sin\theta_{\rm W}\\
          \sin\theta_{\rm W} & \cos\theta_{\rm W}\\
         \end{pmatrix}
         \begin{pmatrix}
           f_{{\rm CMB}}\\
          f_{\rm e} \\
         \end{pmatrix}\,,
\end{align}
where $\theta_{\rm W}$ denotes the mixing (or Weinberg) angle, and field $f$ stands for either the fundamental field strength tensor $F_{\mu\nu}$ or the effective gauge field $a_\mu$ in unitary-Coulomb gauge \cite{bookHofmann}. An effective, propagating external field $a^3_{\rm \mu,{\rm e}}$ thus couples to the droplet charge as 
\eab
\label{eele}
q=\bar{g}_{\rm e}(\lambda_0)\cos\theta_{\rm W}(\lambda_0)&=&-\frac{27}{64}q_{\rm e}(\lambda_0)\cos\theta_{\rm W}(\lambda_0)\nonumber\\ &=&-\frac{27}{64}\frac{4\pi}{e(\lambda_0)}\cos\theta_{\rm W}(\lambda_0)\,,
\eae
compare with Eqs.\,(\ref{meanvalueMPxi}) and (\ref{mixthetaw}). Note that we continue to define the dimensionless temperature $\lambda$ as $\lambda\equiv\frac{2\pi T}{\Lambda_{\rm e}}$ where $\Lambda_{\rm e}$ 
denotes the Yang-Mills scale of SU(2)$_{\rm e}$, amounting to a few keV \citep{ThierryHofmann2022}. The precise value under 
gauge-theory mixing is computed in Sec.\,\ref{massformula}. Also, we now denote by $\lambda_0$ the temperature where the 
pressure of the {\sl mixing plasmas} of SU(2)$_{\rm CMB}$ and SU(2)$_{\rm e}$ vanishes.
Likewise, an effective, external field $a^3_{\rm \mu,{\rm CMB}}$ would couple to a reduced 
monopole charge
\eqb
\label{cmbele}
q_{\rm CMB}(\lambda_0)\sin\theta_{\rm W}(\lambda_0)=\frac{4\pi}{e(\kappa^{-1}\lambda_0)}\sin\theta_{\rm W}(\lambda_0)\,.
\eqe
Since SU(2)$_{\rm CMB}$ is in the deconfining phase inside and outside the droplet, there is no mirror-charge factor. Also, due to the small value of the Yang-Mills scale $\Lambda_{\rm CMB}$ the directed propagation of $a^3_{\mu,{\rm CMB}}$ is constrained to long wavelengths \citep{Hofmann:2015oia2,bookHofmann}.
Here, we introduce  
\eqb
\label{defkappa}
\kappa\equiv{\Lambda_{\rm CMB}}/{\Lambda_{\rm e}}\ll 1\,, 
\eqe
based on $\Lambda_{\rm CMB}\sim 10^{-4}\,$eV. Note that $e(\kappa^{-1}\lambda_0)=\sqrt{8}\pi$ to a very good approximation. 
In the droplet's bulk, however, (thermal) gauge fields $a^3_{\rm \mu,{\rm e}}$ and $a^3_{\mu,{\rm CMB}}$ couple with reduced strength as 
\eqb
\label{monoinbl_e}
q_{\rm e}(\lambda_0)\cos\theta_{\rm W}(\lambda_0)=\frac{4\pi}{e(\lambda_0)}\cos\theta_{\rm W}(\lambda_0)
\eqe
and as in Eq.\,(\ref{cmbele}) to their respective monopoles.
As in the SM \citep{Weinberg:1967tq,Glashow:1959wxa,Salam:1964ry}, the universality of the two reduced couplings in Eqs.\,(\ref{monoinbl_e}) and (\ref{cmbele}) (mixed gauge field couples to each charge with the same strength) thus requires 
\eqb
\label{chargeaftermixing}
q_{\rm CMB}(\lambda_0)\sin\theta_{\rm W}(\lambda_0)=q_{\rm e}(\lambda_0)\cos\theta_{\rm W}(\lambda_0)\,.
\eqe
Eq.\,(\ref{chargeaftermixing}) is the definition exploited in Sec.\,\ref{WA} to thermodynamically determine the value of $\theta_{\rm W}$. Eq.\,(\ref{chargeaftermixing}) also implies the following correspondences between the Cartan algebra in SU(2)$_{\rm CMB}\times$SU(2)$_{\rm e}$ and the Cartan algebra (unitary gauge) in the gauge group of the Standard Model SU(2)$_{\rm W}\times$U(1)$_Y$ at $\lambda_0$:
\eab
\label{cart}
&&{\rm Cartan (\mathfrak{su}(2)}_{\rm e})={\rm u(1)}_Y\,,\nonumber\\ 
&&{\rm Cartan(\mathfrak{su}(2)}_{\rm CMB})={\rm Cartan (\mathfrak{su}(2)}_{\rm W})\,.
\eae
Considering that SU(2)$_{\rm CMB}$ is deeply in its deconfining phase close to the 
droplet surface at $\lambda_{c^\prime}$, keeping the quasiparticle masses of its off-Cartan fluctuations at values well below $10^{-4}\,$eV, see also the discussion in Sec.\,6, we are led to make the following assignment between off-Cartan members of the algebras of SU(2)$_{\rm CMB}\times$SU(2)$_{\rm e}$ and SU(2)$_{\rm W}\times$U(1)$_Y$ at $\lambda_{c^\prime}$, however:
\eqb
\label{offcart}
{\rm off\mbox{-}Cartan (\mathfrak{su}(2)}_{\rm e})={\rm off\mbox{-}Cartan (\mathfrak{su}(2)}_{\rm W})\,.
\eqe
That is, in the SU(2)$_{\rm CMB}\times$SU(2)$_{\rm e}$ model $\theta_{\rm W}$ may change as a process relegates its focus from the deconfining bulk (definition of droplet's charge in interaction with long-wavelength external probes) to the boundary shell 
(electroweak conversions from heavier to lighter charged leptons). 

In the SM's (extremely successful) nonthermal, weak-coupling approach and complete (rather than stepwise) SU(2) gauge-symmetry breaking, invoked by a fundamentally charged Higgs-field of (unitary-gauge) neutral vacuum expectation $v_{\rm H}=246\,$GeV, the Weinberg angle $\theta_{\rm W}$ can also be defined via 
\eqb
\label{massratioweinberg}
\cos\theta_{\rm W}=\frac{m_{W^\pm}}{m_{Z_0}}\,,
\eqe
where $m_{W^\pm}$ and $m_{Z_0}$ denote the masses of 
the vector bosons mediating the weak force. In the SU(2)$_{\rm CMB}\times$SU(2)$_{\rm e}$ model proposed here, the large hierarchy of 
measured vector-boson mass of 80.4\,GeV ($W^\pm$) and 90.2\,GeV ($Z^0$) characterises the decoupling of effective gauge fields in SU(2)$_{\rm e}$ 
at the deconfining-preconfining ($W^\pm$) and preconfining-confining ($Z^0$) phase boundaries as a consequence of divergent effective coupling constants. Collectively, this physics takes place within the thick boundary shell. The effective Higgs mechanisms at play are an adjoint one in the former and an Abelian one in the latter situation. Therefore, although the effective {\sl nonthermal} Cartan mode of SU(2)$_{\rm e}$, $a^3_{\mu,{\rm e}}$, is modelled by the gauge group U(1)$_Y$ in the Standard Model, the 
effective off-Cartan modes of SU(2)$_{\rm e}$, $a^{1,2}_{\mu,{\rm e}}$, close to the Hagedorn transition play the role of the massive vector bosons $W^{\pm}_\mu$ of SU(2)$_{\rm W}$ in the Standard Model. Also, the effective dual, massive U(1) mode of the preconfining phase in SU(2)$_{\rm e}$ plays the role of the massive vector boson $Z_0$ in the Standard Model. Due to phase mixture in SU(2)$_{\rm e}$, setting in slightly below $T_{c, \rm e}$, the masses $m_{W^\pm}$ and $m_{Z_0}$ cannot be defined thermodynamically in the SU(2)$_{\rm CMB}\times$SU(2)$_{\rm e}$ model. 
Because of this and since, thermodynamically, all massive gauge bosons are solely generated in SU(2)$_{\rm e}$ a definition of $\theta_{\rm W}$ via Eq.\,(\ref{massratioweinberg}) would be meaningless, and we have to resort to Eq.\,(\ref{chargeaftermixing}) for a 
useful thermodynamical definition of $\theta_{\rm W}$. 
In the absence of CP-violating terms in the fundamental Yang-Mills action, like in the SM, there is no theoretical basis in the SU(2)$_{\rm CMB}\times$SU(2)$_{\rm e}$ model for why only left-handed charged currents couple to the weak, massive gauge fields.
Also, lepton universality, which comprises instable mu- and tau-leptons and their neutrinos, is a concept difficult to describe thermodynamically by an extension of our present model to SU(2)$_{\rm CMB}\times$SU(2)$_{\rm e}\times$SU(2)$_{\rm \mu}\times$SU(2)$_{\rm \tau}$. 
However, certain considerations are made in Sec.\,6.
Finally, it is not clear how a neutral scalar excitation of mass 126\,GeV -- the Higgs boson -- collectively emerges in the SU(2)$_{\rm CMB}\times$SU(2)$_{\rm e}$ model as an excitation of the phase- and gauge-group mixed plasma of the thick boundary shell. The Standard Model is highly efficient and successful in addressing all these features. For the time being, we therefore must confine our discussion of electroweak physics to the electromagnetic coupling of electrons to thermal photons or propagating electromagnetic waves. Still, such limited understanding of the underpinning of electroweak physics in the SU(2)$_{\rm CMB}\times$SU(2)$_{\rm e}$ model produces values of $\alpha$ and $\theta_{\rm W}$ which are close to their SM values, as we shall see in Secs.\,\ref{Value of alpha} and \ref{WA}, respectively.

\subsection{Mixing angle \texorpdfstring{$\theta_{\rm W}$}{0W}
\label{WA}}

During the creation of an electron-positron pair, we must assume that the fundamental gauge fields that initiate the formation of the two droplets are purely $A_{\mu,{\rm e}}$ since the initial two-photon state is nonthermal ($\theta_{\rm W}=0$) \citep{bookHofmann}. Once the droplet volumes define themselves and after internal thermal equilibrium is attained, there is a fixed mixing angle $\theta_{\rm W}$. An intriguing feature of the Standard Model is that the value of 
$\theta_{\rm W}$ can be computed at a certain four-momentum transfer from the value of $\theta_{\rm W}$ measured at another four-momentum transfer. 
The intriguing feature of SU(2) Yang-Mills thermodynamics is that the value of $\theta_{\rm W}$ at zero four-momentum transfer appears to be computable independently of what is assumed so far in deriving the value of $\alpha$. Let us demonstrate this.

Due to mixing of SU(2)$_{\rm CMB}$ and SU(2)$_{\rm e}$ the deconfining-phase bulk pressure of the droplet $P_{\rm bulk}$ at temperature 
$T\ge T_{c, \rm e}$ ($T_{c, \rm e}$ the critical temperature for the deconfining-preconfining phase transition in SU(2)$_{\rm e}$) and to one-loop accuracy is given as\footnote{Components of the 
perfect-fluid thermal energy-momentum tensor $\theta_{\mu\nu}$ such as the pressure $P$ or the energy density $\rho$ are bilinear functionals of 
the fundamental field-strength tensor $F_{\mu\nu}$. Therefore, mixing coefficients $\sin\theta_{\rm W}$ or $\cos\theta_{\rm W}$ appear in squared form.}
\eab
\label{Pbulk}
P_{\rm bulk}(T)&=&\left(1-\sin^2\theta_{\rm W}(T)\right)\,P_{\rm e}(T)+\sin^2\theta_{\rm W}(T) P_{\rm CMB}(T)\nonumber\\ 
&=&P_{\rm e}(T)+\sin^2\theta_{\rm W}(T)(P_{\rm CMB}(T)-P_{\rm e}(T))\nonumber\\ 
&=&P_{\rm e}(T)+\sin^2\theta_{\rm W}(T)\left(P_{\rm CMB,gs}(T)-P_{{\rm e},{\rm gs}}(T)\right)\nonumber\\ 
& &+\,\sin^2\theta_{\rm W}(T)\left(P_{\rm CMB,3\,pols}(T)-P_{{\rm e},{\rm 3\,pols}}(T)\right)\,,\nonumber\\
\eae
where $P_{\rm e}, P_{\rm CMB}$ denote the total pressures in deconfining SU(2)$_{\rm e}$ and SU(2)$_{\rm CMB}$, respectively. 
They are defined in detail in Eqs.\,(\ref{eandcmbP}), (\ref{defdimlessrho2}), and (\ref{defdimlessrhoperPOl2}) below. The indices 'gs' and '3\,pols' refer to the contributions to these pressures arising from the respective ground states and the 
two effective gauge-mode excitations with three polarisations (due to the adjoint Higgs mechanisms invoked by the thermal ground states). 
The contributions of the effective gauge-mode excitations with two polarisations (massless modes) cancel exactly between SU(2)$_{\rm e}$ and SU(2)$_{\rm CMB}$ in the term on the right-hand side of in Eq.\,(\ref{Pbulk}) which is proportional to $\sin^2\theta_{\rm W}(T)$. Note that the pressure of a monopole intrinsically is nil. 

From now on, $T_0$ and $T_{0,\rm e}$ are agreed to denote the zeros of $P_{\rm bulk}$ and $P_{\rm e}$, respectively. Therefore, Eq.\,(\ref{Pbulk}) implies that 
\eqb
\label{Tversch}
T_0=\frac{\lambda_0}{2\pi}\Lambda_{\rm e}\le T_{0,\rm e}=\frac{\lambda_{0,{\rm e}}}{2\pi}\Lambda_{\rm e}\ \ \ \ (\pi>\theta_{\rm W}\ge 0)\,.
\eqe 
 The inequality (\ref{Tversch}) holds since the differences $P_{\rm CMB,gs}-P_{{\rm e},{\rm gs}}$ and $P_{\rm CMB,3\,pols}-P_{{\rm e},{\rm 3\,pols}}$ both are positive due to Eq.\,(\ref{defkappa}) as well as \footnote{We identify: 
 \eab
 P_{\rm CMB,3\,pols}(T)&=&-12\frac{\kappa^4(\Lambda_{\rm e}\lambda)^4}{(2\pi)^6}\tilde{P}(2a(\kappa^{-1}\lambda))\,,\nonumber\\ 
 P_{\rm CMB,gs}(T)&=&-2\kappa^3\Lambda_{\rm e}^4\lambda\,,\nonumber\\ 
 P_{\rm e,3\,pols}(T)&=&-12\frac{(\Lambda_{\rm e}\lambda)^4}{(2\pi)^6}\tilde{P}(2a(\lambda))\,,\nonumber\\ 
 P_{\rm e,gs}(T)&=&-2\Lambda_{\rm e}^4\lambda\,.
 \eae}
 Eqs.\,(\ref{eandcmbP}), (\ref{defdimlessrho2}), and (\ref{defdimlessrhoperPOl2}). This means that $P_{\rm e}(T_0)$ must be negative which, in turn, implies that $T_0\le T_{0,\rm e}$. According to \citep{Hofmann:2017lmu}, one has 
\eqb
\label{0ldeT}
T_{0,{\rm e}}=1.32\,T_{c,\rm e}\ \ \mbox{or} \ \ \ \lambda_{0,\rm e}=1.32\,\lambda_{c,\rm e}=18.31\,.
\eqe 
Because of Eqs.\,(\ref{Tversch}), (\ref{defkappa}) and (\ref{0ldeT}), we may to a very good approximation use the asymptotic value 
\eqb
\label{qasymptmix}
q_{\rm CMB}=\frac{4\pi}{e(\kappa^{-1}\lambda_0)}\approx \sqrt{2}\,
\eqe
in Eq.\,(\ref{chargeaftermixing}). Thus, we have  
\eqb
\label{tantheta}
\tan\theta_{\rm W}(\lambda_0)=\frac{e(\kappa^{-1}\lambda_0)}{e(\lambda_0)}\approx \frac{\sqrt{8}\pi}{e(\lambda_0)}\,.
\eqe 
Substituting Eq.\,(\ref{tantheta}) in Eq.\,(\ref{Pbulk}), we finally arrive at
\eab
\label{Pbulkee}
0&=&P_{\rm bulk}(T_0)\nonumber\\ 
&=& \left(1-\sin^2\theta_{\rm W}(\lambda_0)\right)\,P_{\rm e}(T_0)+\sin^2\theta_{\rm W}(\lambda_0) P_{\rm CMB}(T_0)\nonumber\\ 
&=&\left(1-\frac{e^2(\kappa^{-1}\lambda_0)}{e^2(\lambda_0)+e^2(\kappa^{-1}\lambda_0)}\right)\,P_{\rm e}(T_0)\nonumber\\ 
& &+\,\frac{e^2(\kappa^{-1}\lambda_0)}
{e^2(\lambda_0)+e^2(\kappa^{-1}\lambda_0)}P_{\rm CMB}(T_0)\nonumber\\ 
&\approx&\left(1-\frac{8\pi^2}{e^2(\lambda_0)+8\pi^2}\right)\,P_{\rm e}(T_0)+\frac{8\pi^2}{e^2(\lambda_0)+8\pi^2}P_{\rm CMB}(T_0)\,.\nonumber\\ 
\eae
Explicitly, the pressures $P_{\rm e}(T_0)$ and $P_{\rm CMB}(T_0)$ are given as \citep{bookHofmann}
\eab
\label{eandcmbP}
P_{\rm e}(T_0)&=&-\Lambda_{\rm e}^4\bar{P}(\lambda_0,a(\lambda_0)) \nonumber\\ 
P_{\rm CMB}(T_0)&=&-\kappa^4\Lambda_{\rm e}^4\bar{P}(\kappa^{-1}\lambda_0,a(\kappa^{-1}\lambda_0))\,,
\eae
where 
\eqb
\label{defdimlessrho2}
\bar{P}(\lambda,a(\lambda))\equiv
\frac{2\lambda^4}{(2\pi)^6}\left[2\tilde{P}(0)+6\tilde{P}(2a)\right]+2\lambda\
\eqe
and 
\eqb
\label{defdimlessrhoperPOl2}
\tilde{P}(y)\equiv \int_0^{\infty} dx\,x^2\,\log\left[1-\exp\left(-\sqrt{x^2+y^2}\right)\right]\ \,.
\eqe
The advantage of writing $P_{\rm CMB}(T_0)$ as in Eq.\,(\ref{eandcmbP}) is that the precise value of 
$\kappa\ll 1$ is not required to be known in order to extract $\lambda_0$ and $e(\lambda_0)$ 
from the condition in Eq.\,(\ref{Pbulkee}). This is due to the rapid vanishing of the quantity $a(\lambda)$ as $\lambda\to\infty$, see Eq.\,(\ref{ae}). Solving Eq.\,(\ref{Pbulkee}) for $\lambda_0$ numerically, we have 
\eqb
\label{NewT}
\lambda_0=16.3=1.18\,\lambda_{c,\rm e}\,, \ \ \ e(\lambda_0)=14.88\,.
\eqe
Indeed, due to mixing the zero $\lambda_0$ of the bulk pressure turns out to be smaller than $\lambda_{0,{\rm e}}$, compare Eqs.\,(\ref{0ldeT}) and (\ref{NewT}). 

Finally, solving Eq.\,(\ref{tantheta}) 
for $\theta_{\rm W} (\lambda_0)$ subject to Eq.\,(\ref{NewT}), we obtain
\eqb
\label{thetaweinnum}
\theta_{\rm W}(\lambda_0)\equiv\theta_{\rm W}(Q=0)=\arctan\frac{\sqrt{8}\pi}{e(\lambda_0)}=30.84^\circ\,,
\eqe
where, with a slight abuse of notation, the argument $Q=0$ indicates that the system is probed at vanishing four-momentum transfer (resolution referring to the maximum of the moduli of the Mandelstam variables $s,t,u$ that contribute to the probing process). This is close to the experimentally obtained value of the Weinberg angle. In 
\citep{MoellerscatteringWA} the value of $\theta_{\rm W}$ was extracted from the parity-violating asymmetry in fixed target electron-electron 
scattering at a resolution of $Q=0.1612$\,GeV as
\eqb
\label{WAes}
\theta_{\rm W}(Q=0.1612\,{\rm GeV})=29.23^\circ \cdots 29.40^\circ\,.
\eqe
The latest particle data group quotation of $\theta_{\rm W}$, measured at $Q$ equal to the mass $m_Z$ of the $Z$-boson, is \citep{PDG22}
\eqb
\label{WAZ}
\theta_{\rm W}(Q=m_Z=90.2\,{\rm GeV})=28.73^\circ \cdots28.75^\circ\,.
\eqe
Note the tendency of a mild increase of $\theta_{\rm W}$ from Eq.\,(\ref{WAZ}) to Eq.\,(\ref{WAes}) for vastly 
decreasing values of resolution (logarithmic running). The thermodynamically determined value of $\theta_{\rm W}$ in Eq.\,(\ref{thetaweinnum}) 
is our prediction for $\theta_{\rm W}(Q=0)$.

The Weinberg angle at $Q=0$ appears to be determined quite accurately from bulk thermodynamics without having to make explicit assumptions on the finite-volume physics of the droplet's thick boundary shell, where SU(2)$_{\rm e}$ and SU(2)$_{\rm CMB}$ and all phases of the former theory mix. Implicitly, we assume, however, that the pressure of this shell as exerted onto the bulk is zero. Note that for an infinite-volume system, where stability constraints are irrelevant, we have $\theta_{\rm W}(\lambda)=45^\circ$ for the mixing of the theories SU(2)$_{\rm e}$ and SU(2)$_{\rm CMB}$ in the conformal limit $\lambda\gg\lambda_0$.

\subsection{Value of \texorpdfstring{$\alpha$}{alpha}}
\label{Value of alpha}

Considering gauge-theory mixing inside the deconfining-phase bulk of the droplet and assuming the external, effective gauge field, which probes the droplet charge, to reside in SU(2)$_{\rm e}$, we may use Eq.\,(\ref{eele}) to obtain 
\eqb
\label{alphaunmixed3}
\alpha=\frac{q^2}{4\pi}=\frac{\left(-\frac{27}{64}q_{\rm e}(\lambda_0)\cos\theta_{\rm W}\right)^2}{4\pi}\,.
\eqe
Appealing to the value of $\theta_{\rm W}$ in Eq.\,(\ref{thetaweinnum}) and to the value of $e(\lambda_0)$ in Eq.\,(\ref{NewT}) yields
\eqb
\label{inverse_alpha}
\alpha^{-1}=134.3\,.
\eqe
This deviates by only 2\,\% from the experimental value 
of Eq.\, (\ref{definition_alpha}). 

However, in contrast to the determination of the Weinberg angle $\theta_{\rm W}$, which is a quantity that depends 
only on the stability of bulk thermodynamics in a finite volume ($P_{\rm bulk}=0$) and on the universality of monopole charges, the determination of the fine-structure constant $\alpha$ hinges in addition on phase mixing {\sl within} SU(2)$_{\rm e}$ and mixing of the gauge groups SU(2)$_{\rm e}$ and SU(2)$_{\rm CMB}$ inside the thick boundary shell. For this boundary shell, defined by a vanishing monopole charge in SU(2)$_{\rm e}$ within the radial extent 
$\bar{r}\le r\le r_0$ ($\bar{r}=\xi r_0$ the mean radius w.r.t. the a priori distribution of monopole location), we also had to assume vanishing pressure. To understand the physics of the thick boundary shell better is the subject of future work. An according improvement of the mirror-charge construction used so far to define the droplet's charge but also the effect of droplet revolution on the probability density $\rho$ in Eq.\,(\ref{rho}) could decrease the difference between the value of $\alpha$ in Eq.\,(\ref{inverse_alpha}) and its experimental value in Eq.\,(\ref{definition_alpha}).  

\subsection{Impact of mixing on mass formula and length-scale hierarchy in powers of \texorpdfstring{$\alpha$}{alpha} }
\label{massformula}

A modelling of the mixing effects within the thick boundary shell 
is beyond the scope of the present article. Therefore, in what follows, we content ourselves with estimating $T_{c, \rm e}$ 
and the droplet radius $r_0$ under gauge-theory mixing, and we assume that the energy densities are the same within the droplet's thick boundary shell and the droplet's bulk.
Let us determine the shift in Yang-Mills scale $\Lambda_{\rm e}$ when changing the model of the free electron based on pure SU(2)$_{\rm e}$ to a model that invokes mixing of SU(2)$_{\rm e}$ and SU(2)$_{\rm CMB}$. For a pure SU(2)$_{\rm e}$ model, the electron's rest-mass $m_{\rm e}$, which coincides with the lowest circular breathing frequency $\omega_0=e(\lambda_{0,\rm e})\,H_\infty(T_0)$ of the monopole \citep{Forg_cs_2004,Fodor_2004,Hofmann:2017lmu}, is given as (re-writing Eq.\,(5) of \citep{Grandou:2018pjp})
\eab 
\label{massfromm0pureE}
m_{\rm e}&=&e(\lambda_{0,\rm e})\,H_\infty(T_0)=m_m(T_0)+\frac{4\pi}{3}r_0^3\rho(T_0)\nonumber\\
&=&H_\infty(T_0)\left(\frac{4\pi}{e(\lambda_{0,\rm e})}+\frac{64\pi}{3\lambda_{0,\rm e}^4}\chi^3\bar{\rho}(\lambda_{0,\rm e})\right)\,,
\eae
where the dimensionless plasma energy density $\rho(T_0)/\Lambda_{\rm e}^4 \equiv \bar{\rho}(\lambda,a(\lambda))$ is defined as 
\eqb
\label{defdimlessrho}
\bar{\rho}(\lambda,a(\lambda))\equiv\frac{2\lambda^4}{(2\pi)^6}\left[2\tilde{\rho}(0)+6\tilde{\rho}(2a)\right]+2\lambda\,,
\eqe
$m_m(T_0)$ denotes the mass of a BPS monopole originating from the dissociation of a (anti)caloron of maximally nontrivial holonomy, and $\chi\equiv r_0 H_\infty(T_0)\equiv \pi r_0 T_0$. The reason why the BPS limit is considered here is that a nontrivial-holonomy caloron sets the value of the asymptotic, adjoint Higgs field of its constituent monopole and antimonopole solely in terms of its holononmy and not dynamically by minimization of potential energy density, see \citep{Lee_1998,Kraan_1998}.
In Eq.\,(\ref{defdimlessrho}), we have introduced 
\eqb
\label{defdimlessrhoperPOl}
\tilde{\rho}(y)\equiv \int_0^{\infty} dx\,x^2\,\frac{\sqrt{x^2+y^2}}{\exp\left(\sqrt{x^2+y^2}\right)-1}\,, 
\eqe
and in Eq.\,(\ref{massfromm0pureE}) the quantity $H_\infty(T_0)=\pi T_0$ refers to the modulus of the (anti)caloron gauge-field component $A_4(r\to\infty)=a_4$, see \citep{Hofmann:2017lmu}. In the case of gauge-theory mixing of SU(2)$_{\rm CMB}$ with SU(2)$_{\rm e}$, we generalise Eq.\,(\ref{massfromm0pureE}) to 
\begin{align}
\label{massfromm0mixing}
\frac{m_{\rm e}}{H_\infty(T_0)}=&\,e(\lambda_0)\cos{\theta_{\rm W}}+e(\kappa^{-1}\lambda_0)\sin{\theta_{\rm W}}\nonumber\\ 
=&\cos{\theta_{\rm W}}\frac{4\pi}{e(\lambda_0)}+\sin{\theta_{\rm W}}\frac{4\pi}{e(\kappa^{-1}\lambda_0)}+\frac{64\pi}{3\lambda_0^4}\chi^3\nonumber\\ 
&\times\,\left[\cos^2{\theta_{\rm W}}\bar{\rho}(\lambda_0,a(\lambda_0))+\kappa^4\sin^2{\theta_{\rm W}}\bar{\rho}\left(\frac{\lambda_0}{\kappa},a\left(\frac{\lambda_0}{\kappa}\right)\right)\right].
\end{align}
Solving the second half of Eq.\,(\ref{massfromm0mixing}) for $\chi$ after appealing to Eqs.\,(\ref{NewT}) 
and (\ref{thetaweinnum}), we obtain 
\eqb
\label{solxi}
r_0={\rm 5.09}\,H^{-1}_\infty(T_0)\,.
\eqe
From Eqs.\,(\ref{solxi}) and (\ref{NewT}) we deduce
\eqb
\label{r0calr}
\frac{r_0}{|\phi|^{-1}(T_0)}=0.155\,,
\eqe
indicating that the droplet is contained deeply within the central region of a typical caloron or anticaloron which contributes to the emergence of the deconfining thermal ground state of SU(2)$_{\rm e}$. Yet, we have $r_0/\beta_0=1.62$ such that the radial integral in the definition of $\phi$'s phase reasonably well represents a sinusoidal $\tau$-dependence, see \citep{bookHofmann}, p.\,127. As a consequence of Eqs.\,(\ref{NewT}), (\ref{solxi}) we have
\eqb
\label{r_0Lambde}
r_0=0.622\,\Lambda_{\rm e}^{-1}\,.
\eqe
With this corrected value of droplet radius $r_0$ due to gauge-theory mixing, compare with Eq.\,(\ref{r0SU2epure}) for the case of pure SU(2)$_{\rm e}$, the ratio of screening length $l_s$ to $r_0$ of Eq.\,(\ref{lsvsr0}) is reduced compared to the case of pure SU(2)$_{\rm e}$ as 
\eqb
\label{redratiolsr0}
\frac{l_s}{r_0}\sim 44.1\,.
\eqe
Still, this is sufficiently larger than unity to justify the mirror-charge construction of Sec.\,\ref{MCC}. Recall, that such a construction requires 
the Coulomb nature of the static U(1) potential of the monopole away from its core region. 
\noindent The first part of Eq.\,(\ref{massfromm0mixing}) yields\\
\eqb
\label{mastoHinfty}
\frac{m_{\rm e}}{H_\infty(T_0)}=17.33\,. 
\eqe
Together with Eq.\,(\ref{solxi}) this implies a ratio of reduced Compton length $r_c$ to $r_0$ as
\eqb
\label{RrCr0}
\frac{r_c}{r_0}=\frac{1}{17.33\times 5.09}=\frac{1}{88.3}\,.
\eqe
That is, due to mixing, the core-size of the trapped monopole, which according to the first part of Eq.\,(\ref{massfromm0pureE}) is close to $r_c$ \cite{Fodor_2004}, becomes even more pointlike compared to the droplet's extent than for the case of a pure SU(2)$_{\rm e}$ model where $r_c/r_0=1/52.40$. Eq.\,(\ref{mastoHinfty}) also implies that  
\eqb
\label{LambdaEm0}
\frac{m_{\rm e}}{\Lambda_{\rm e}}=141.82\,
\eqe
or $\Lambda_{\rm e}=3.60\,$keV or $T_{c,{\rm e}}=7.95\,$keV. 
\begin{figure}[H]
\centering
\includegraphics[width=\columnwidth]
{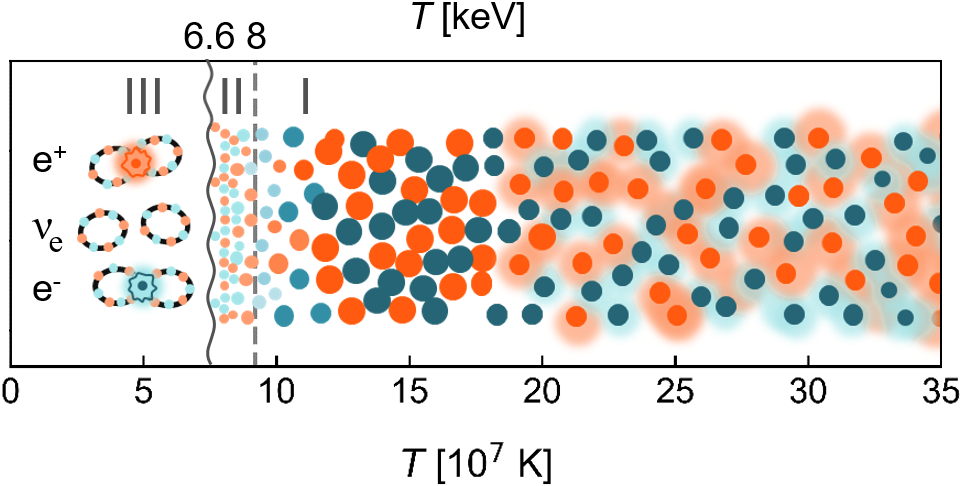}
\caption{Phase diagram of SU(2)$_{\rm e}$ with $T_{c,{\rm e}}=0.923\times 10^8\,$K (7.95\,keV) in the infinite-volume limit. There are three distinct phases: (I) the deconfining phase, (II) the preconfining phase, and (III) the confining phase.}
\label{Fig:PhaseDiagramE}
\end{figure}

According to Eq.\,(\ref{RrCr0}), the ratio of droplet radius $r_0$ to Bohr radius $a_0$ is
\eqb
\protect\label{dropletBohr}
\frac{r_0}{a_0}=\frac{r_0}{r_c}\frac{r_c}{a_0}=88.3\frac{r_c}{a_0}=88.3\,\alpha=0.64\,.
\eqe
Such a large ratio of order unity is in line with Louis de Broglie's proposal \cite{DeBroglieRev} that the free electron at rest represents the oscillation of a thermal plasma of finite extent and spatially constant amplitude driven by a vibrating monopole. Namely, within the deconfining bulk of the droplet at rest, the superposition of spherical wavelets of constant frequency and wavelength 
but highly random phase and variable origin plausibly yield a plasma oscillation of nearly constant spatial amplitude — the basic insight of de Broglie's deduction of the electron's matter wavelength paving the way for wave mechanics \cite{Schrodinger:1926xyk}. Note that also outside the droplet, the spatially decaying Coulomb field $a^3_4$ keeps oscillating at the droplet's frequency $\omega_0=m_{\rm e}$. This also is true of the energy density or the pressure of the plasma component due to SU(2)$_{\rm CMB}$ at temperatures that locally are higher than the CMB baseline temperature within a certain, local environment of the droplet.

\section{Discussion and Summary}

In this work we have proposed the electron to be a figure-eight shaped soliton formed by a thin center-vortex loop in SU(2) Yang-Mills theory where the region of selfintersection is an extended droplet. This soliton is stable and immersed into the confining ground state of an SU(2) Yang-Mills theory of scale 3.60\,keV. 
Moreover, the soliton is subject to the confining ground states of other Yang-Mills theories with higher scales and the Cosmic Microwave Background (CMB). In the present work, we associate the CMB with the deconfining phase of an SU(2) Yang-Mills theory of scale $\sim 10^{-4}$\,eV (SU(2)$_{\rm CMB}$) \citep{bookHofmann,Hofmann:2009yh}. 

Based on the definition of the Bohr magneton, we also link the electron's magnetic moment to revolutions of the droplet's effective charge. These revolutions are induced by the unresolved monopoles and antimonopoles moving along and constituting the center-vortex loop. The contribution of the thin vortex lines to the total mass of this soliton is negligible, see \citep{bookHofmann}. However, vortex lines may play a role in producing nonlocal interactions that magnetically correlate electrons in a 2D plane \citep{Moosmann:2008fr}. \\

Within the bulk region of the droplet an electric monopole is trapped which internally vibrates due to kicks issued by the quantum physics of the thermal ground state \citep{bookHofmann}. This ground state locally superimposes a caloron and an anticaloron center and overlapping (anti)caloron peripheries. Under the selfconsistent assumption that the droplet's bulk can be represented by infinite-volume thermodynamics, we have derived the following statements: 

(i) Bulk stability, that is, the vanishing of the thermodynamical pressure deep inside the droplet implies a mixing angle between the Yang-Mills theories SU(2)$_{\rm e}$ and SU(2)$_{\rm CMB}$ at one and the same temperature $T_0$ which practically coincides with the value of the weak mixing (or Weinberg) angle of the Standard Model of Particle Physics. 

(ii) The value of the electromagnetic fine-structure constant $\alpha$ at vanishing energy-momentum transfer is reasonably well approximated by a mirror charge construction, suggested  by the high electric conductivity of the droplet's thick boundary shell. In addition, the derivation of the value of $\alpha$ relies on (a) a thermodynamical SU(2)$_{\rm e}$ and SU(2)$_{\rm CMB}$ gauge-theory mixing plus monopole charge universality, (b) stability of the thick boundary shell by its assumed, vanishing mean pressure, and (c) the statistical average of the droplet's charge over the bulk volume w.r.t. a spatially homogeneous distribution, cut off at the mean radius (condition that the thick boundary shell is excluded as a region for  positioning the isolated monopole of the SU(2)$_{\rm e}$ component of the bulk plasma). 

(iii) The droplet's mass formula, in generalisation of \citep{ThierryHofmann2022} now considering  SU(2)$_{\rm e}$ and SU(2)$_{\rm CMB}$ gauge-theory mixing, predicts a droplet size $r_0$ which is comparable with the Bohr radius $a_0$. For the core size of the monopole $r_c$ (the reduced Compton wave length \citep{Fodor_2004,Forg_cs_2004,DeBroglieRev}) we therefore obtain $r_c\sim\alpha\,r_0\sim \alpha\,a_0=\alpha^{-1}\,r_{\rm e}$ where $r_{\rm e}$ denotes the classical electron radius. As a consequence, the monopole core 
indeed is an (unresolved) point particle on the scales of the droplet's extent and on the radius $|\phi|^{-1}$ of a (anti)caloron which constitutes the thermal ground state. 

(iv) The Yang-Mills scale $\Lambda_{\rm e}$ of SU(2)$_{\rm e}$ and the critical temperature $T_{c,{\rm e}}$ are derived from this mass formula, applying it to the mass of the electron $m_{\rm e}=511\,$keV. One has: $\Lambda_{\rm e}=3.60\,$keV or $T_{c,{\rm e}}=7.95\,$keV. This implies a critical temperature $T_{c^\prime,{\rm e}}$ of the Hagedorn transition as $T_{c^\prime,{\rm e}}=\frac{11.57}{13.87}\,7.95\,$keV$=6.63\,$keV. When the spatio-temporal design of fusion plasmas is optimised to accomplish steady-state operation, involving comparable electron temperatures (and high electron densities), these results should be taken into account in devising magnetic and inertial plasma confinement strategies. 

(v) There is an environment to the stationary droplet of an extent yet to be specified which is characterised by a plasma of deconfining SU(2)$_{\rm CMB}$ phase, slightly hotter than the CMB. This deconfining plasma vibrates at a frequency $\sim m_{\rm e}$. The monopole of SU(2)$_{\rm CMB}$ may alternate its location from the droplet's bulk region into this environment and 
vice versa.\\ 

Let us now briefly discuss possible links of these results to phenomena discussed in the literature. 
In our present approach, we would interpret the particle at rest as the droplet whose vibrating (standing) external Coulomb potential is transformed into a propagating wave upon boost \citep{DeBroglieFP}, giving rise to observable (self-)interference effects to statistically determine the droplet's position \citep{1976AmJPh..44..306M}. 
As mentioned in the introduction, we do expect a considerable impact of the thermodynamical approach to the electron proposed here in better understanding collective plasma phenomena at electron temperatures starting at around $T\sim T_c=6.6$\,keV in experiments with magnetic plasma confinement which were not predicted by conventional magneto-hydrodynamics. These could include the formation of an edge transport barrier associated with a pressure pedestal, edge-localised modes, magnetic instabilities, and ion-orbit losses in the high-confinement mode \citep{LAGGNER2019479,Stacey}.
Note that the electron plasma density $n_e$ in conventional tokamaks and stellerators is $n_e \sim 10^{20}\,{\rm m}^{-3}$ while a macroscopically stabilised plasma droplet at $T=T_0=9.38$\,keV with an energy density of $\rho(T_0)=1.77\times 10^4$\,keV$^4$ represents a number density of percolated electrons of $n_e \sim 1.79\times 10^{28}\,{\rm m}^{-3}$. Therefore, eight orders of magnitude in electron density are missing in order to achieve a macroscopically stabilised plasma state. Still, the above-mentioned effects at much lower electron densities may point to the here-proposed model of the electron.\\

\begin{figure*}
\centering
\onecolumn
\protect\caption{
Creation and decay of a muon: 
\textbf{a)} A not yet quantum equilibrated muon droplet, defined as the selfintersection region of a center-vortex loop, is created within the confining phase of SU(2)$_{\mu}$. This droplet contains three types of 
monopoles: $q_\mu$, $q_{\rm e}$, $q_{\rm g}$.
\textbf{b)} Energy in surplus to the muon quantum mass is dissipated into the surroundings. The monopoles $q_{\rm e}$, $q_{\rm g}$ are not trapped by the thick boundary shell of the muon droplet (turquoise) and are free to leave or re-enter. 
A single center-vortex loop of SU(2)$_{\rm e}$ is fused with $q_{\rm e}$ to form a figure-eight shaped object subject to a new electron droplet. The monopole $q_{\rm g}$ (dark blue) in SU(2)$_{\rm CMB}$ outside the electron droplet, at the CMB's present temperature, would be reduced to a massless and zero-charge point particle \protect\cite{Hofmann:2009yh}. 
\textbf{c)} The size of the electron droplet (blue) is comparable with the Bohr radius $a_0$ and contains the muon droplet. 
\textbf{d)} The muon droplet is dissolved by the thick boundary shell of the electron droplet, rendering $q_\mu$ an irrelevant, massless, and zero-charge point particle which casts a figure-eight shaped center-vortex loop into a single one: the muon neutrino $\nu_\mu$. The localised energy injected to the boundary shell thus is a transient process which can be interpreted as a $W^{-}$ boson in the Standard Model of Particle Physics.
\textbf{e)} The final states of muon decay are the electron and $\nu_\mu$.
\textbf{f)} Feynman-diagram of muon decay in the Standard Model of Particle Physics (the Feynman diagram was taken from \protect\cite{FeynmanPicture}).
}
\twocolumn
\includegraphics[width=18cm]{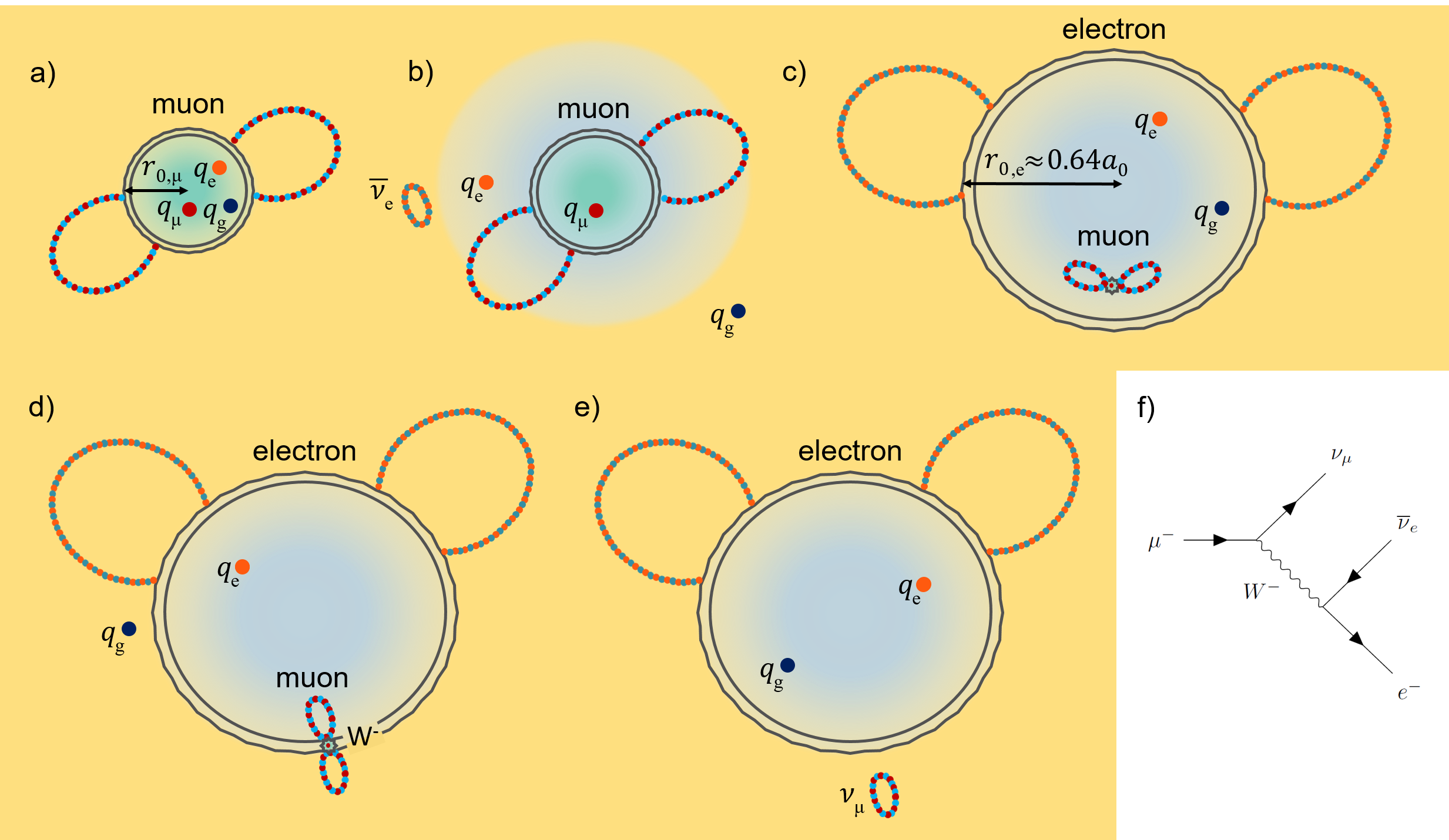}
\label{Fig:muon_decay}
\end{figure*}

An important question, which arises due to a modelling of the electroweak parameters $\theta_{\rm W}$ and $\alpha$ that apparently is particular to the electron as a thermal and stable quantum particle, concerns these parameters' experimentally enshrined universality across the electroweak interactions of all leptons. As for the unstable, charged leptons $\mu$ and $\tau$ the reason why their charge is identical to that of the electron would be as follows. As soon as a $\tau$ or a $\mu$ lepton is created at rest, the according droplets in SU(2)$_{\rm CMB}\times$SU(2)$_{\rm e}\times$SU(2)$_{\mu}$, see Fig.\,\ref{Fig:muon_decay}\,a), and in SU(2)$_{\rm CMB}\times$SU(2)$_{\rm e}\times$SU(2)$_{\mu}\times$SU(2)$_{\tau}$, respectively, disperse the energies invested in surplus to their quantum mass, defined by a generalisation of Eq.\,(\ref{massfromm0mixing}), into their surroundings and can only trap an SU(2)$_{\mu}$ or an SU(2)$_{\tau}$ monopole, respectively, see Fig.\,\ref{Fig:muon_decay}\,b) for the muon droplet. The other monopoles are free to leave or re-enter this droplet. In case of SU(2)$_{\rm CMB}$ this ab initio dispersion of energy does not define new phase boundaries since the CMB represents the deconfining phase (likely very close to the deconfining-preconfining transition \cite{bookHofmann}). In case of SU(2)$_{\mu}$ or SU(2)$_{\rm e}$ new droplets are formed that embed the initial droplet, see Fig.\,\ref{Fig:muon_decay}\,c) for the case of SU(2)$_{\rm e}$. This process of (cascading) droplet formation invokes formerly single, round-point center-vortex loops from confining-phase ground states to define the respective droplets as their stabilised regions of selfintersection: single center-vortex loops thus turn into figure-eight shaped center-vortex loops. In the Standard Model of Particle Physics, this subprocess refers to the absorption of an antineutrino, compare  Fig.\,\ref{Fig:muon_decay}\,b) and Fig.\,\ref{Fig:muon_decay}\,f). After quantum equilibration, the initial droplets' bulks exhibit defined mixing angles for four or three deconfining SU(2) Yang-Mills theories by two or three charge-universality conditions (in general: in the droplet representing the $N$th lepton family there are $N$ charge-universality conditions fixing the $N$ independent components of a unit vector in an $(N+1)$-dimensional Euclidean space or $N$ mixing angles), respectively, and a defined temperature by the condition of vanishing bulk pressure. Since the according leptons are unstable, no (temporally) coherently propagating waves in the Cartan algeba's of SU(2)$_{\tau}$ and SU(2)$_{\mu}$ exist that could externally probe these charges and the magnetic moments that relate to the center-vortex loops extending from $\mu$- and $\tau$-droplets. Therefore, even though the electron droplet contains $\mu$- and $\tau$-droplets and their respective, trapped monopoles, only the trapped monopole and the magnetic moment provided by the center-vortex loop in SU(2)$_{\rm e}$ are seen externally. For the charge of $\mu$ and $\tau$ leptons, we thus are back at the derivation of the charge of an electron (or fine-structure constant $\alpha$), see Sec.\,\ref{Value of alpha}. Their decays can be figured as processes, where embedded droplets of much smaller radii $r_{0,\mu}\sim\frac{m_{0,{\rm e}}}{m_{0,\mu}}r_{0,{\rm e}}$ and $r_{0,\tau}\sim\frac{m_{0,{\rm e}}}{m_{0,\tau}}r_{0,{\rm e}}$  ($m_{0,i}$, $r_{0,i}$ the rest mass, droplet radius of charged lepton $i$ with $i={\rm e},\mu,\tau$), subject to gauge theory mixing in SU(2)$_{\rm CMB}\times$SU(2)$_{\rm e}\times$ SU(2)$_{\mu}$ and SU(2)$_{\rm CMB}\times$SU(2)$_{\rm e}\times$SU(2)$_{\mu}\times$SU(2)$_{\tau}$, respectively, by eventual contact with the thick boundary shell of an embedding droplet (for the $\tau$ lepton droplet these embedding droplets are the droplets of SU(2)$_{\mu}$ and SU(2)$_{\rm e}$, for the $\mu$ lepton droplet this is the droplet of SU(2)$_{\rm e}$) dissolve to feed their high energy densities and massive, trapped monopoles, locally into those of the nonthermally distorted respective boundary shells. This local investment of energy into the boundary shell can be thought of as a transient excitation of a $W^\pm$-boson in the Standard Model of Particle Physics. As a result, an energetic neutrino is emitted: a figure-eight shaped center-vortex loop looses its droplet together with the trapped monopole when interacting with the thick boundary shell to transform into a single center-vortex, see Fig.\,\ref{Fig:muon_decay}\,d) for $\mu$-decay and  Fig.\,\ref{Fig:muon_decay}\,f) the according subprocess in the Standard Model of Particle Physics.\\ 

To relate hadrons to pure Yang-Mills thermodynamics is much 
harder than for leptons. Hadrons are complex quantum systems of confined 
(anti)quarks that are effectively and efficiently described by Quantum Chromodynamics \cite{PhysRevLett.30.1343,PhysRevLett.30.1346}. To address the emergence of (anti)quarks as electrically fractionally charged particles within pure Yang-Mills theories of one and the same electric-magnetic parity is impossible. Rather, a derivation of quark properties probably would require an interplay and mixing of electric-magnetic dual SU(3) Yang-Mills models to allow a version of the fractional Quantum Hall effect \cite{PhysRevLett.48.1559,PhysRevB.23.5632,PhysRevLett.50.1395} to take place.\\

In closing, we would like to state clearly that there cannot be any doubt that the Standard Model of Particle Physics represents a milestone development in accurately and efficiently describing the interactions of leptons and hadrons. Essentially, this theory rests on well organised weak-coupling expansions that implement the gauge principle in a perturbatively consistent way \cite{Taylor:1971ff,Slavnov:1972fg} and that are applicable to any so far probed energy-momentum transfer. 
A thermodynamical approach to the {\sl interactions} of leptons and hadrons in terms of pure Yang-Mills theory is inferior to the Standard Model. 
What the Standard Model is incapable of delivering though is a ground-state structure doing justice to cosmological observations \cite{Riess_1998,Perlmutter_1999}, to provide a useful framework for thermal and nonthermal phase transitions \cite{TurbPlasma}, a postdiction of the absolute values of (some of) its dimensionless parameters, and a deeper grasp of the nature of particle-matter-wave duality concerning charged leptons. As the present work intended to demonstrate, a computation of two of the Standard Model's dimensionless parameter values appears to be feasible. The here-proposed venue is still far from addressing other in-built features of the Standard Model such as parity violation of the weak interactions, a derivation of the electron's magnetic moment including the anomalous quantum behaviour, a quantitative grasp of the entries of the CKM matrix, or the fractional electric charges of quarks. We hope to gain more insight into these problems in the future. 

\section{Data availability}
\noindent  
The python notebook which has been used for the calculations in this paper is available on \href{https://git.uni-wuppertal.de/meinert/electroweak-parameters-from-yang-mills-thermodynamics}{\hspace{0.1mm}\includegraphics[scale=0.05]{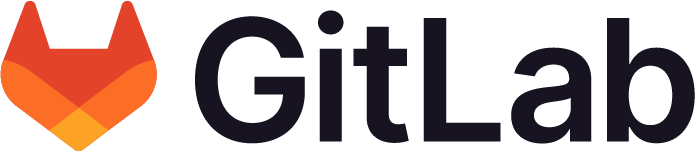}}

\section{Acknowledgements}
RH acknowledges helpful discussion with Manfried Faber and Thierry Grandou, and he wishes to thank his wife Karin Thier for her patience and enduring support. JM thanks Karl-Heinz Kampert and Alexander Sandrock for insightful discussions and useful comments. 
\noindent JM's work is supported by the Vector Foundation under grant number P2021-0102 and partially by SFB 1491. 

\bibliographystyle{mnras}
\bibliography{BibSU2CMB}
\label{lastpage}
\end{document}